\input epsf
\input harvmac
\noblackbox
\newcount\figno
\figno=0
\def\fig#1#2#3{
\par\begingroup\parindent=0pt\leftskip=1cm\rightskip=1cm\parindent=0pt
\baselineskip=11pt
\global\advance\figno by 1
\midinsert
\epsfxsize=#3
\centerline{\epsfbox{#2}}
\vskip 12pt
\centerline{{\bf Figure \the\figno:} #1}\par
\endinsert\endgroup\par}
\def\figlabel#1{\xdef#1{\the\figno}}

\font\cmss=cmss10
\font\cmsss=cmss10 at 7pt
\def\rlx{\relax\leavevmode}
\def\inbar{\vrule height1.5ex width.4pt depth0pt}
\def\IC{\relax\,\hbox{$\inbar\kern-.3em{\rm C}$}}
\def\IN{\relax{\rm I\kern-.18em N}}
\def\IP{\relax{\rm I\kern-.18em P}}
\def\ZZ{\rlx\leavevmode\ifmmode\mathchoice{\hbox{\cmss Z\kern-.4em Z}}
 {\hbox{\cmss Z\kern-.4em Z}}{\lower.9pt\hbox{\cmsss Z\kern-.36em Z}}
 {\lower1.2pt\hbox{\cmsss Z\kern-.36em Z}}\else{\cmss Z\kern-.4em Z}\fi}
\def\narrowplus{\kern -.04truein + \kern -.03truein}
\def\narrowminus{- \kern -.04truein}
\def\narrowminussub{\kern -.02truein - \kern -.01truein}

\def\YM{Yang-Mills}

\def\ker{{\rm Ker}}

\def\cl{\centerline}

\def\vac{\vert 0\rangle}
\def\lvac{\langle 0 \vert}
\def\O{{\cal O}}

\def\b{{\beta}}
\def\a{{\alpha}}
\def\g{{\gamma}}
\def\e{{\epsilon}}
\def\vol{{\rm vol}}
\def\Pf{{\rm Pf}}
\def\l{{\lambda}}
\def\w{{\omega}}
\def\p{{\psi}}
\def\r{{\rightarrow}}
\def\cp{{\overline \psi}}
\def\frac#1#2{{#1\over #2}}

%
%
\def\eqnn#1{\xdef #1{(\secsym\the\meqno)}\writedef{#1\leftbracket#1}%
\global\advance\meqno by1\wrlabeL#1}
\def\eqna#1{\xdef #1##1{\hbox{$(\secsym\the\meqno##1)$}}
\writedef{#1\numbersign1\leftbracket#1{\numbersign1}}%
\global\advance\meqno by1\wrlabeL{#1$\{\}$}}
\def\eqn#1#2{\xdef #1{(\secsym\the\meqno)}\writedef{#1\leftbracket#1}%
\global\advance\meqno by1$$#2\eqno#1\eqlabeL#1$$}

\lref\rK{N. Ishibashi, H. Kawai, Y. Kitazawa and A. Tsuchiya, hep-th/9612115.}
\lref\rCallias{C. Callias, Commun. Math. Phys. {\bf 62} (1978), 213.}
\lref\rPD{J. Polchinski,  Phys. Rev. Lett. {\bf 75} (1995) 47.}
\lref\rWDB{E. Witten,  Nucl. Phys. {\bf B460} (1996) 335.}
\lref\rSSZ{S. Sethi, M. Stern, and E. Zaslow, Nucl. Phys. {\bf B457} (1995)
484.}
\lref\rGH{J. Gauntlett and J. Harvey, Nucl. Phys. {\bf B463} 287. }
\lref\rAS{A. Sen, Phys. Rev. {\bf D53} (1996) 2874; Phys. Rev. {\bf D54} (1996)
2964.}
\lref\rWI{E. Witten, Nucl. Phys. {\bf B202} (1982) 253.}
\lref\rPKT{P. K. Townsend, Phys. Lett. {\bf B350} (1995) 184.}
\lref\rWSD{E. Witten, Nucl. Phys. {\bf B443} (1995) 85.}
\lref\rASS{A. Strominger, Nucl. Phys. {\bf B451} (1995) 96.}
\lref\rBSV{M. Bershadsky, V. Sadov, and C. Vafa, Nucl. Phys. {\bf B463}
(1996) 420.}
\lref\rBSS{L. Brink, J. H. Schwarz and J. Scherk, Nucl. Phys. {\bf B121}
(1977) 77.}
\lref\rSS{S. Sethi, unpublished.}
\lref\rCH{M. Claudson and M. Halpern, Nucl. Phys. {\bf B250} (1985) 689.}
\lref\rSM{B. Simon, Ann. Phys. {\bf 146} (1983), 209.}
\lref\rGJ{J. Glimm and A. Jaffe, {\sl Quantum Physics, A Functional Integral
Point of View},
Springer-Verlag (New York), 1981.}
\lref\rADD{ U. H. Danielsson, G. Ferretti, B. Sundborg, Int. J. Mod. Phys. {\bf
A11} (1996) 5463\semi   D. Kabat and P. Pouliot, Phys. Rev. Lett. {\bf 77}
(1996), 1004.}
\lref\rDKPS{ M. R. Douglas, D. Kabat, P. Pouliot and S. Shenker,
Nucl. Phys. {\bf B485} (1997), 85.}
\lref\rhmon{S. Sethi and M. Stern, Phys. Lett. {\bf B398} (1997), 47.}
\lref\rBFSS{T. Banks, W. Fischler, S. H. Shenker, and L. Susskind,
Phys. Rev. {\bf D55} (1997) 5112.}
\lref\rBHN{ B. de Wit, J. Hoppe and H. Nicolai, Nucl. Phys. {\bf B305}
(1988), 545\semi
B. de Wit, M. M. Luscher, and H. Nicolai, Nucl. Phys. {\bf B320} (1989),
135\semi
B. de Wit, V. Marquard, and H. Nicolai, Comm. Math. Phys. {\bf 128} (1990),
39.}
\lref\rT{ P. Townsend, Phys. Lett. {\bf B373} (1996) 68.}
\lref\rLS{L. Susskind, hep-th/9704080.}
\lref\rFH{J. Frohlich and J. Hoppe, hep-th/9701119.}
\lref\rAg{S. Agmon, {\it Lectures on Exponential Decay of Solutions of
Second-Order Elliptic Equations}, Princeton University Press (Princeton) 1982.}
\lref\rY{P. Yi, hep-th/9704098.}
\lref\rDLhet{ D. Lowe, hep-th/9704041.}

\Title{\vbox{\hbox{hep--th/9705046}\hbox{IASSNS--HEP--97/45, DUK-M-97/5
}}}
{D-Brane Bound States Redux\footnote{$^*$}{A preliminary version of this paper
was circulated informally in November, 1996.}}

\smallskip
\centerline{Savdeep Sethi\footnote{$^1$} {sethi@sns.ias.edu} }
\medskip\centerline{\it School of Natural Sciences}
\centerline{\it Institute for Advanced Study}\centerline{\it
Princeton, NJ
08540, USA}
\vskip 0.12in
\centerline{and}
\vskip 0.12in
\centerline{Mark Stern\footnote{$^2$} {stern@math.duke.edu} }
\medskip\centerline{\it Department of Mathematics}
\centerline{\it Duke University}\centerline{\it Durham, NC 27706, USA}


\vskip .2in

We study the existence of D-brane bound states at threshold in Type II
string theories. In a
number of situations, we can reduce the question of existence to quadrature,
and the study of
a particular limit of the propagator for the system of D-branes. This involves
a derivation of an index theorem for a family of non-Fredholm operators. In
support of the
conjectured relation between compactified eleven-dimensional supergravity
and Type IIA string
theory, we show that a bound state exists for two coincident zero-branes.
This result also provides support for the conjectured description of M-theory
as a matrix
model.  In
addition, we provide further evidence that there are no BPS bound states for
two and
three-branes twice wrapped on Calabi-Yau vanishing cycles.
\vskip 0.1in
\Date{4/97}

\newsec{Introduction}

Remarkable progress by Polchinski in describing the solitons of Type II
string theory has
provided the means by which many conjectured dualities involving string
theories and M-theory
can be stringently tested \rPD. The low-energy
dynamics of coincident D-branes has been described by Witten \rWDB, who
reduced the question
of finding BPS bound states to one
of studying the vacuum structure of various supersymmetric Yang-Mills
theories.  In simple
cases, the BPS mass formula forbids the decay of a charged particle
saturating the mass bound;
hence, ensuring stability.  However, there are a number of situations in
which a particle is
required that is only marginally stable against decay.  Showing the
existence of such
particles, with energies at the decay threshold, is the goal of this paper.
A similiar
problem arose for finite $ SU(2)$ N=2 Yang-Mills theory, studied in \rSSZ\
and \rGH, where
certain dyon bound states at threshold were shown to exist.  The situations
we shall presently
study are significantly more difficult because the Hamiltonians are not as well
behaved, and gauge
invariance provides an added complexity.

Let us briefly recall the low-energy dynamics of coincident Dirichlet $
p$-branes, described
in \rWDB. The world volume theory is the dimensional reduction of
ten-dimensional N=1
Yang-Mills to the $ p+1$-dimensional world volume of the brane. For a single
brane, the gauge
theory is abelian, and the dynamics therefore trivial in the infrared. For $
N$ coincident
branes, the gauge symmetry is enhanced to $ U(N)$ rather than $ U(1)^N$.
After factoring out a
$ U(1)$ corresponding to the center of mass motion, the existence of a bound
state requires
that  the  remaining $ SU(N)$ $ p+1$-dimensional Yang-Mills theory possess a
normalizable
supersymmetric vacuum. The bosonic potential for this model generally has
flat directions, and
so we encounter the problem of bound states at threshold. If a bound state is
required by a conjectured duality, there is a consistency check, described by
Sen \rAS, that can sometimes be performed. In favorable cases, one might be
able
to further compactify one direction of the superstring
theory. If a bound state exists prior to compactification, it should give rise
to BPS states in the further compactified theory which, for appropriate choices
of momentum along the circle, are no longer marginally bound. The existence of
these states can then be analyzed with more conventional techniques. Of course,
for this consistency check, there have to be enough remaining uncompactified
directions so that problems with infra-red divergences do not arise. More
generally,
however, the
question of bound states at threshold must be addressed. Note that a
normalizable state for a
theory in a compact space generally does not remain normalizable when the
volume is taken to
infinity.  The spectrum can and often does change discontinously, and showing
the existence of the bound state in the non-compact
situation requires a separate analysis.

In a similar spirit, we can arrive at descriptions of the effective dynamics
of $p$-branes
multiply-wrapped on supersymmetric cycles of a compactification space. In
the case of
$p$-branes wrapped on $p$-cycles, the resulting description of the
low-energy dynamics is some flavor of
quantum mechanics, although not generally just a supersymmetric gauge
theory. Our aim in this
paper is to address the fundamental issue -- the existence of flat
directions in the potential
-- which arises in studying binding in these situations. This analysis
generalizes the
discussion in \rhmon, where we argued for the existence of a marginal bound
state of a zero-brane and a four-brane, to the case where the gauge group is
non-abelian. We shall see that
there are very subtle issues that arise as a result of this
complication.

The most exciting reason for studying this question is, however, the
remarkable
conjecture that M-theory may be described in terms of zero-brane dynamics in
the limit where
the number of branes goes to infinity \rBFSS. This conjecture is, in part,
founded on previous
work studying the relation between supermembranes, and the $N \rightarrow
\infty$ limit of
type IIA zero-brane quantum mechanics \refs{\rBHN,\rT}. In order for the
M(atrix) model to have a chance at describing M theory, we need to be able to
find states in the quantum mechanics which correspond to the gravitons of
eleven-dimensional supergravity. The bound state that we shall find is
precisely
one of these particles. In the process of showing that such a bound state
exists, we will provide a detailed study of the behavior of the propagator for
the two zero-brane system when the zero-branes are far apart. There
are a number of complications that make this analysis quite subtle. During the
lengthy course of our investigation (which pre-dates the M(atrix) model), a
number of germane papers have appeared. Among these papers have been
interesting discussions of zero-brane scattering in various approximations
\refs{\rADD, \rDKPS},
and more recently, an exciting extension of the original matrix model
conjecture
to the case of finite $N$ \rLS. There has also been an explicit argument
showing that there are no normalizable ground states in a particular simplified
matrix model \rFH, a heuristic attempt to argue for the existence of zero-brane
bound states \rDLhet, and a recent paper which has some overlap with our
results
\rY.

In the following section, we consider the case of $p=0$. We describe a seven
parameter family
of theories, which are the primary focus of this discussion. This family of
theories is
derived fundamentally from the quantum mechanics describing the zero-brane
in type IIA string
theory by adding mass deformations. These parameters allow us to `flow down'
from ten dimensions to models that correspond to the reduction of N=1
Yang-Mills
in lower dimensions by taking various mass terms to infinity. We discuss
general
features of these models, including the various
physical scenarios
in which they arise. Our approach to the question of counting bound states
is described in
section three. There, we argue that the $L^2$ index for
this class of
supersymmetric quantum mechanical Yang-Mills theories is actually computable.
This involves a discussion of $L^2$ index theory for non-Fredholm operators,
which is an area of mathematics that is relatively unexplored.  In section
four, we study the question of
two-particle binding in these models, and we derive a formula for the principal
contribution to the index. The final section is a study of the two-particle
propagator in the limit where the two particles are far apart. With this
analysis, we can compute a subtle additional contribution to the index. The way
in which this contribution arises involves some rather surprising
cancellations. In the class of models that we investigate, we find
that only the case which corresponds to the reduction of supersymmetric
Yang-Mills from ten dimensions can have a unique bound state. This answers, in
large part, the question of why the large $N$ limit of the reduced
ten-dimensional Yang-Mills theory should be distinguished from the large $N$
limit of reductions of lower-dimensional Yang-Mills theories.

\newsec{Quantum Mechanical Gauge Theory}
\subsec{General Comments}

Let us begin by considering models that arise from reducing  supersymmetric
d+1-dimensional $
SU(N)$ Yang-Mills to quantum mechanics; see, for instance, \rCH\ for the first
discussion of quantum mechanical gauge theories, or perhaps \rBSS.
Whether the
Yang-Mills theory contains additional matter multiplets does not
significantly change the
following discussion; so, for simplicity, we shall assume no additional
matter. On reducing
the connection $ A_\mu$, we obtain scalar coordinates $ x^i$ where $
i=1,...,{\rm d}$ which
take values in the adjoint representation of the gauge group. We introduce
canonical momenta
obeying,
$$ [ x^i_A, p^j_B] = i \delta_{AB} \delta^{ij}, $$
where the subscript $ A$ is a gauge index. With the generators $ T^A$ for
the adjoint
representation normalized so that $ \Tr (T^A T^B) = N \delta^{AB}$, the
Hamiltonian for the
system takes the general form,
\eqn\hamilt{H = {1\over 2N} \Tr (p^i p^i) + V(x) + H_F.}
The bosonic potential $ V(x)$ is polynomial in $ x$, and generally has flat
directions. The
term $ H_F$ is quadratic in the fermions and linear in $ x$. Specific
examples will be studied
in the following subsection.  The $ A_0$ equation of motion gives a set of
constraints, $
C_A$, which must vanish on physical states by Gauss' law. The constraints
obey the algebra,
\eqn\constraints{ [C_A, C_B ] = i f_{ABC} C_C, }
where $ f_{ABC}$ are the structure constants. The constraints further obey
the commutation
relations $ [C_A, H] = [C_A, Q]=0,$ where $ Q$ is a supersymmetry generator.
 The
supersymmetry algebra closes on the Hamiltonian if the constraints are set
to zero. An
$N$-particle BPS bound state corresponds to a normalizable, gauge-invariant
ground state for
this supersymmetric system.

Without detailed computation, what might we infer about the structure of the
ground state?
Away from the flat points, the wave function for the ground state will decay
exponentially.
The only interesting asymptotic behavior is expected near points where the
potential is small.
A preliminary comment about the structure of the flat directions is in
order: for gauge group
$ SU(N)$, there are $ d_c = (d-1)(N-1) + (N^2-1)$ commuting
directions around a flat point, and $d_a = (d-1)(N^2 - N)$ non-commuting
directions.  Let us
consider the structure of the potential in the neighborhood of a flat point.
As we shall
subsequently describe in detail, the potential can be approximated by $V
\sim - {1\over 2} r^2
|v|^2$, where $ v$ parametrizes the transverse directions, and $ r$ is a
radial coordinate for
the flat directions.  The Hamiltonian is then essentially a set of bosonic
and fermionic
harmonic oscillators for the transverse directions, and a free Laplacian
along the flat
directions. The frequency for oscillation along the massive directions
depends on $ r$. This observation provides one way of seeing that there are
no scattering
states in the spectrum of the bosonic Hamiltonian for these models, as
discussed in \rSM.
Somewhat surprisingly, the spectrum of the bosonic models only contains
discrete states. To
construct a scattering state along the flat direction, one would want to put
the transverse
harmonic oscillators into their ground states; however, the zero point
energy of the
oscillators increases with $ r$, essentially forbidding finite energy
scattering states. The
same argument does not  apply to the supersymmetric case, since the ground
state energy for
the
additional fermions now cancels the zero point energy from the bosons, as
required by
supersymmetry. If this were not the case, the subtleties in counting
zero-brane bound states
would not exist!

In a first approximation for large $ r$, any zero-energy wavefunction, $\psi
(x)$, roughly takes
a product form corresponding to placing the transverse  oscillators into the
ground state,
$\psi(x) \sim  g(r, \theta) e^{-r|v|^2/2}$, where $\theta$ are angular
variables for the flat directions. The leading dependence of $ g(r, \theta)$ on
$r$ is believed to be  power
law decay for
large $ r$. Acting with the Hamiltonian for the massive directions on
this wavefunction
yields zero, since the zero point energies of the bosons and fermions
cancel. We can now
explain the key difficulty in studying the approximate asymptotic wavefunction:
can the decay exponent be accurately estimated?

We note that this issue is critical, and cannot be resolved by
simple
approximations of the asymptotic behavior. For instance, even in this
approximation, the function $
g(r, \theta)$ is not simply
the solution of a free Laplacian for the $ d_c$-dimensional space of flat
directions since the
Laplacian, which for the radial coordinate is given by,
\eqn\free{ \Delta_r =  - {1\over
r^{d_c-1} }
{{\partial \over \partial r}} r^{d_c-1} {\partial \over \partial r}, }
also acts on the
harmonic oscillator component of the wavefunction. Actually, it is unlikely
that
the decay exponent can be accurately estimated without at least including the
first excited mode for the massive direction into the approximation.  We should
also note that showing that the decay is fast enough to ensure normalizability
is only a first step toward showing that a bound state exists. The structure of
the wavefunction would need to be studied at small $r$ where the non-abelian
degrees of freedom are important. Currently, the only practical approach is to
develop an appropriate index theory for the problem. As a final comment, note
that the power law behavior of
the asymptotic
ground state wavefunction is a consequence of the lack of a mass gap in the
spectrum. The
supersymmetric theory contains a continuum of states which descend to zero
energy, thanks to
the existence of the flat direction \rBHN.

\subsec{A Family of Models}

We now turn to the models of primary interest to us. Let us recall that
strongly coupled
Type IIA string theory in ten dimensions has a conjectured dual description
as weakly coupled
eleven-dimensional supergravity compactified on an $ S^1$ \refs{\rPKT,\rWSD}.
To
match the
Kaluza-Klein spectrum of  the compactified supergravity theory, Type IIA
string theory
requires electrically charged particle states. The Dirichlet zero-branes,
which carry RR
charge, seem to be the only candidates.  Since there is a single
Kaluza-Klein mode for each
choice of momentum along the circle direction, we desire a single D-brane
bound state for each
$ N$. Proving this conjecture was our original motivation for studying these
theories.

Actually, Sen has argued in \rAS\ that if a unique bound state exists in the
quantum mechanics describing $N$ zero-branes, then the spectrum of ultra-short
multiplets in the toroidally compactified type II string agrees with the
spectrum predicted by U-duality. The  world-volume theory for the D-particle is
given by the dimensional
reduction of  N=1
$9+1$-dimensional Yang-Mills to quantum mechanics.  A Majorana-Weyl spinor
in $9+1$ dimensions
has $ 16$ real components, which means that the resulting quantum mechanical
theory has N=16
supersymmetry. Let $ \gamma^i_{\alpha\beta} $ be a real representation of
the $ SO(9)$
Clifford algebra with $ i=1,...,9$ and $ \alpha = 1,...,16$. These Clifford
matrices satisfy
$$ \left\{ \gamma^i, \gamma^j \right\} = 2 \delta^{ij}. $$
After reduction, the Hamiltonian for this system takes the form,

\eqn\onehamilt{  H  = {1\over 2N} \Tr (p^i p^i) - {1\over 4N} \sum_{ij} \Tr
([ x^i, x^j]^2) -
{1\over 2N} \Tr ( \psi \gamma^i [x^i, \psi ]) , }
where the real fermions $ \psi_{A\a} $ obey:
\eqn\fermions{  \left\{\psi_{A\alpha}, \psi_{B\beta}\right\} = \delta_{AB}
\delta_{\alpha\beta}. }
The Hilbert space is then composed of spinors on which the quantized fermions
act as
elements of a Clifford algebra. The spinor wavefunctions contain an extremely
large
number of
components, even for small $N$, which makes an explicit construction of the
zero
energy bound
state wavefunction at
best difficult.\foot{However, the existence of the nice $Spin(9)$ flavor
symmetry might bring an
explicit construction of the ground state wavefunction within the realm of
possibility. We leave the attempt to construct the explicit solution to braver
souls.} The
supersymmetry algebra takes the form,
\eqn\onesusyalg{ \left\{ Q_{\alpha}, Q_{\beta} \right\} = 2
\delta^{\alpha\beta} H + 2
\gamma^i_{\alpha\beta} x^i_A C_A, }
where,
$$ Q_\a = {1\over N} \g^i_{\a\b} \Tr (\psi_\b p^i) - {i\over 4 N} \Tr (
[\g^i,\g^j] \psi [x^i,
x^j])_\a , $$
while the constraint,
$$ C = - i [x^i, p^i] - {1\over 2} [\psi_\a, \psi_\a], $$
or explicitly,
\eqn\oneconstr{ C_A =  f_{ABC} (x^i_B p^i_C - {i\over 2} \psi_{B
\alpha}\psi_{C\alpha} ). }
The constraint takes exactly the form assumed in the previous discussion. It
is natural to
call this a nine-dimensional model, although it is quantum mechanics, since
there are nine bosonic variables
in the adjoint of
$SU(N)$, and the model is the reduction of a ten-dimensional theory. The flavor
symmetry is clearly $Spin(9)$. Note that there is a nice correlation between
fermion number and flavor representation that is worth mentioning at this
point. The correlation essentially follows from spin-statistics in ten
dimensions: fermionic states in the Hilbert space transform under spinor
representations of the flavor group, while bosonic states appear in
representations of $SO(9)$. If the ground state is unique, it must therefore be
bosonic. There are similar relations for the other models that we shall soon
discuss.

Let us
consider what sort of deformations are possible in this
theory. We would like
to add mass terms to compactify some of the bosonic variables and
effectively reduce the
dimension, but we will also require that the supersymmetry algebra maintain
its nice
structure. In particular, we shall not consider deformations which introduce
additional terms
into the right hand side of the supersymmetry algebra \onesusyalg, which are
linear in
momenta. The mass deformations that we shall describe correspond, in special
cases, to
breaking N=4 Yang-Mills in four-dimensions to N=2 or N=1 by giving masses to
various chiral
fields in the adjoint representation, and reducing the corresponding model
to quantum
mechanics.

To describe the allowed deformations, choose a real supersymmetry generator,
$Q=Q_\a$. The
generator can be split into terms involving momenta, and terms independent
of momenta. Those
depending on momenta can be expressed, schematically, as $ \l^i_A p^i_A$,
where $\l^i$ is a
real fermion, and $i$ runs from $1$ to $9$. This leaves us with seven real
fermions, $\w^j_A$ in the adjoint of the gauge group,
unpaired with a momentum operator, but each appearing in $Q$ paired with
an operator, $
f^j_A$, quadratic in the coordinates. The supersymmetry generator is then
roughly,
$$ Q \sim \l^i_A p^i_A + \w^j_A f^j_A + \ldots.$$
The seven fermions, $\w^j$, then represent our
deformation degrees
of freedom. We can add any reasonable operator to $f^j$, independent of the
momenta, and not
generate a new term linear in momenta in the expression for $\{Q, Q \}.$ There
are many interesting possible deformations that preserve at least one
supersymmetry. Some deformations can give quite exotic classical minima of the
resulting bosonic potential. This is a topic that merits further investigation.
As a special prosaic case,
we could add the perturbation $m x^i$ to one of the $f^j$, which would lift
some of the
flat directions.
This is the family of deformations to which we shall restrict our discussion.
More
explicitly, consider a
term $f^j$ which squares to give the term in the potential, $ |f^j|^2$.
Adding
the term $m x^i$
to $f^j$ changes the potential to $|f^j+ mx^i|^2 $. Taking
$m\rightarrow \infty$
then effectively decouples $x^i$ from the model. In this way, we generate a
seven parameter
family of models which depend on the values of the allowed masses for seven
of the
coordinates. Note that taking all masses to infinity leaves us with a
two-dimensional model,
and further compactification is not possible without introducing additional
terms linear in
the momenta into the supersymmetry algebra.

There are two cases of particular interest: the three and five-dimensional
models. These
models correspond to the reduction of N=1 Yang-Mills from four and six
dimensions,
respectively. For completeness and to fix annoying normalizations, we shall
describe the
Hamiltonian and supersymmetry algebra for both models explicitly.

In the three-dimensional case, the Hamiltonian is given by,
\eqn\twohamilt{  H  = {1\over 2N} \Tr (p^i p^i) - {1\over 4N} \sum_{ij} \Tr
([ x^i, x^j]^2) +
{1\over N} \Tr ({ \overline \psi} \sigma^i [x^i, \psi ]) , }
where the index $ i=1,2,3.$ The $ \sigma^i$ are the Pauli matrices, and the
complex fermions $
\psi$ obey the anti-commutation relations:
$$  \left\{\psi_{A\alpha},\overline{\psi}_{B\beta} \right\} = \delta_{AB}
\delta_{\alpha\beta}, $$
where $\a=1,2$. The supersymmetry generators are now complex, but still take a
form similar to the previous example,
$$ Q_\a = \sigma^i_{\a\b} \p_{A\b} p^i_A  - {1\over 4} f_{ABC} [ \sigma^i,
\sigma^j ]_{\a\b} \p_{A\b} x^i_B x^j_C, $$
while the constraints are given by,
\eqn\constrtwo{  C_A = f_{ABC} (x^i_B p^i_C - i \cp_{B\a}\p_{C\a} ). }
The supersymmetry algebra is now,
\eqn\twosusy{ \eqalign{ \left\{ Q_{\alpha}, Q_{\beta} \right\} & = 0, \cr
\left\{ {\overline Q}_{\alpha}, {\overline Q}_{\beta} \right\} & = 0, \cr
\left\{ {\overline Q}_{\alpha}, Q_{\beta} \right\} & = 2 \delta_{\a\b} H - 2
\sigma^i_{\a\b} x^i_A C_A.}}
The most glaring difference between this model and the nine-dimensional
zero-brane case is
that the Hilbert space is now a Fock space with a canonical vacuum. This model
is quite special because the Pauli matrices form a Lie algebra, and so the
complex supercharge can be expressed as \rCH,
$$ Q_a = \sigma^i_{\a\b} \p_{\b} (p^i_A - {i\over 2} f_{ABC} \e^{ijk} x^j_B
x^k_C ). $$
After introducing a potential, $W = {1\over 6} f_{ABC} \e^{ijk} x^i_A x^j_B
x^k_C$, we can conjugate the supercharge in the following way:
$$ e^{W} Q_a e^{-W} = \sigma^i_{\a\b} \p_{\b} p^i_A.$$
The study of the ground state wavefunctions then takes on a cohomological
flavor since the supercharge acts roughly as the operator,
$$ Q \sim d + dW\wedge, $$
on the wavefunctions, which we can view as differential forms. We expect that,
in this case, there should then be an explicit proof from studying the spectrum
directly that shows there are no zero-energy $L^2$ wavefunctions for this
model.

The five-dimensional case is governed by the Hamiltonian:
\eqn\threehamilt{  H  = {1\over 2N} \Tr (p^i p^i) - {1\over 4N} \sum_{ij}
\Tr ([ x^i, x^j]^2)
+ {1\over N} \Tr ({ \overline \psi} \g^i [x^i, \psi ]). }
The index $ i$ now runs from $ 1$ to $ 5$, and the matrices $ \g$ are
elements of the $ SO(5)$
Clifford algebra. Again, the fermions are complex, and obey the relations,
$$  \left\{\psi_{A\alpha}, {\overline \psi}_{B\beta} \right\} = \delta_{AB}
\delta_{\alpha\beta}, $$
where $ \a = 1,\ldots ,4$. The constraint has a form identical to the
previous case \constrtwo, and the supersymmetry algebra is given by:

\eqn\threesusy{ \eqalign{ \left\{ Q_{\alpha}, Q_{\beta} \right\} & = 0, \cr
\left\{ {\overline Q}_{\alpha}, {\overline Q}_{\beta} \right\} & = 0, \cr
\left\{ {\overline Q}_{\alpha}, Q_{\beta} \right\} & = 2 \delta_{\a\b} H - 2
\gamma^i_{\a\b} x^i_A C_A.}}

\subsec{Wrapped D-branes}

Some of the models described in the previous section have already been
realized from wrapped
D-brane configurations. Let us begin by considering type IIB
string theory, and the case of  three-branes wrapped on a collapsing
three-cycle. In his study
of singularities near conifold points of Calabi-Yau manifolds, Strominger
required only a
single massless BPS state wrapped on the vanishing cycle \rASS.   That there
should be no bound
states has been argued from a somewhat different approach in \rBSV. The
geometry of interest
is $ R \times S^3$, where the $ S^3$ shrinks to zero size. Clearly, the
effective theory on $
R \times S^3$ is not N=4 \YM; such a theory would make little sense. Rather
the world-volume
theory of a D-brane on a curved space should be described by a topologically
twisted theory \refs{\rSS, \rBSV}. As the size of the sphere shrinks, only the
light degrees of freedom are relevant. The question, in this situation, then
concerns the
existence of a ground
state in the theory obtained from the dimensional reduction  of
four-dimensional N=1 \YM, as
first mentioned in \rWDB.

We can also check the situation for type IIA, where we perform the same
analysis for the case
of two-branes wrapped on a vanishing two-cycle. The situation is exactly
analogous to the case
described above.  The geometry is now $ R \times S^2$, where the $ S^2$
shrinks to zero size.
The only difference involves the number of supersymmetries. The effective
theory is now the
reduction of N=1 Yang-Mills from six dimensions. Both models were explicitly
described in the
previous subsection. It seems plausible that other D-brane configurations
will realize many,
if not all, of the remaining models which we have discussed.

\subsec{Gauge Invariance}

There is a rather nice feature of some of the computations that we shall
describe that
deserves a separate comment. Whether it provides a hint at how to
 formulate covariantly
M-theory as a matrix model we leave to the judgement of the reader.\foot{When
this section was originally written, the preceding comment seemed most
appropriate. Subsequently, there has been an interesting proposal for a
non-perturbative definition of the type IIB string, given in \rK. The high
temperature limit of the partition function that we describe in this section
reduces precisely to the model in that proposal. The relation between M(atrix)
theory and the proposal in \rK\ seems to be in the spirit of a `T-duality' in
the time direction.}  The
gauge-fields in a
quantum mechanical gauge theory are non-dynamical. They serve only to
enforce the constraint
that all states in the Hilbert space be gauge-invariant. How do we enforce
such a constraint
in the operator formulation? For very high temperatures, the partition
function,
\eqn\partition{ {\rm Z}( \b )  = \int{dx \, \tr \, e^{-\b H} (x,x)} }
can be well-approximated by perturbation theory. The notation that we will use
throughout the paper may be unfamiliar, and so deserves a comment: we will
often
consider traces of some operator, say ${\O}$, which we will denote as,
$$ \tr \, \O (x,y), $$
where by $(x,y)$ we mean the usual propagation of a particle from point $y$ to
point $x$. In an explicit basis of eigenfunctions, $\p_n(x)$, with eigenvalue
$\l_n$ for $\O$, this expression takes the familiar form,
$$ \sum_n \l_n \p_n(x) \p_n(y), $$
where $n$ may index a continuous parameter.

However, in computing the partition function \partition, it is inconvenient
to try to trace
over the gauge invariant spectrum of the Hamiltonian i.e. states $ |\psi
(x)>$ satisfying $C_A  |\psi (x)> = 0$. Our first task is then to implement the
projection onto gauge invariant states
explicitly, so we
can trace over the full, unconstrained spectrum.  The gauge constraints, $
C_A$, split into
two sets of $ SU(N)$ generators: one generates rotations of the $ x^i$,
which we shall denote
$C^b$,  while the other, $C^f$,
generates rotations of the fermions. Let us denote the operator generating a
finite gauge
transformation $ g(t)$ on the fermions by $ \Pi (g(t))$ where we shall drop
the explicit
dependence on $ t$. To project onto gauge invariant states, we insert:
\eqn\projection{\eqalign{
 {\rm Z}( \b ) &=\int_{SU(N)} dt {\int{ dx \, \tr \, e^{i t_A C_A}    \,
e^{-\b H}
(x,x),}} \cr
 &= \int_{SU(N)} dt {\int{ dx \, \tr \, \Pi (g) \,  e^{-\b H} (gx,x)}} ,\cr
}}
where the measure for the $ SU(N)$ integration is chosen so that $
\int_{SU(N)}dt = 1$. The
trace is now over the full Hilbert space, including gauge-variant states.

For small $\b$, we can now construct a reasonable approximation for the
propagator,
\eqn\heat{ e^{-\b H} (x,y) = {1 \over (2\pi \b)^{l/2} } e^{-{|x-y|^2 \over 2
\b}}e^{-\b V}
e^{-\b H_F} + \dots ,}
where $ l = d (N^2-1)$ is the dimension of the space of scalars. We shall
describe this
approximation in somewhat more detail in the following section.  The fermion
projection operator can be expressed as $ \Pi (g(t)) = e^{it_A C^f_A}$,
which yields the
expression,

$$ {\rm Z}( \b )  = \int_{SU(N)} dt \int{dx \, \tr \, {1 \over (2\pi
\b)^{l/2} } e^{-{|x-gx|^2
\over 2 \b}} e^{-\b V} e^{-\b H_F} e^{it_A C^f_A} + \dots  }. $$

As $ \b \rightarrow 0$, we see that the contribution from group elements
away from the
identity element is strongly suppressed. Indeed, we can then replace $ g$ by
$ I + i
\vec{t} \cdot\vec{C}^b$, and the exponential term involving $g$ becomes,
$$ e^{ - | i \vec{t} \cdot\vec{C}^b x|^2 \over 2 \b}.$$
The term $i\vec{t} \cdot\vec{C}^b x$ is more transparent when written as $
{1\over N} \Tr [t,x^i]^2$, but
this is precisely the form of a term in the potential energy, $V$. Indeed,
in this limit, the
gauge parameters combine exactly with the remaining coordinates to give a
trace which is
$SO(d+1)$ symmetric, rather than $SO(d)$ symmetric. Even the fermion
projection operator
combines naturally with $H_F$ to give a complete symmetry between $x^i$ and
$t$, in the
computation of this trace. We shall put this symmetry to good use in
subsequent computations.
Note that for the case of zero-branes in type IIA, the partition function
appears to arise
from a manifestly $SO(10)$ invariant Hamiltonian, without any hint of gauge
constraints.

\newsec{Counting Ground States}

\subsec{Defining the Index}

Ideally, to count the number of normalizable ground states for these models, we
would like to compute the low temperature limit of the partition function,
$$ \int{dx \,\lim_{\b\rightarrow \infty} \tr \, e^{-\b H} (x,x)}. $$
Except for very simple systems, that computation is beyond reach. As usual, we
are then interested in counting the number of $ L^2$ ground states
weighted by $
(-1)^F$ where $ F$ is the fermion number. Therefore, we wish to compute the
index,
\eqn\inddef{\eqalign{
 {\rm Ind}&= \int{ dx \, \lim_{\b \rightarrow  \infty} \, \tr \, (-1)^F
e^{-\b H} (x,x),} \cr
 &=n_B - n_F,\cr
}}
where the trace is over the gauge invariant spectrum of the Hamiltonian. Let us
first note that the index is perfectly well-defined. The only way that the
index \inddef\ could not be counting the net number of ground states is if the
Hamiltonian had an extremely pathological low-energy spectrum, i.e. if the
density of states diverged badly as $E\rightarrow 0$. That is certainly not the
case for the models we are studying.

Whether the index is computable is another question entirely. The purpose of
this section is to argue that our approach to computing the index actually
counts the number of ground states. Before discussing the issues that arise in
the non-Fredholm cases, let us discuss in some detail the situation where there
is a gap in the spectrum. First, when the spectrum is actually discrete, the
twisted partition function is $\b$-independent. In these cases, we can compute
the index in the $\b\rightarrow 0$ limit, which reduces to a perturbative
computation. The $\b\rightarrow 0$ limit is what we will call the principal
contribution to the index. Even in the case where the spectrum is discrete, the
principal contribution, which is often computed as an integral over the
coordinates, $x$, can be shifted to a boundary term. To see this, note that we
should first perform all our analysis on a ball $B_R$, where $|x|<R$, and then
take a limit $R\rightarrow\infty$. We can then write,
$$ \eqalign{ {\rm Ind} &= \lim_{R\r\infty} \int_{|x|<R}{ dx \, \lim_{\b
\rightarrow \infty} \, \tr \, (-1)^F e^{-\b H} (x,x), }\cr
&= \lim_{R\r\infty} \lim_{\b_o\r 0}  \int_{|x|<R}{ dx \, \{ \, \tr \, (-1)^F
e^{-\b_o H} (x,x), + \int_{\b_o}^\infty d\b {\partial \over \partial\b} \, \tr
\, (-1)^F e^{-\b H} (x,x) \} }. \cr }$$
Now, computing $ \partial_\b$ of $\tr \, (-1)^F e^{-\b H}$ brings down $H$,
which we can replace by $Q^2$. When we try to run $Q$ around the trace,
$$ \eqalign{ \tr \, (-1)^F Q^2 e^{-\b H} &= - \tr\, Q (-1)^F Q e^{-\b H} \cr
&= - \tr\, (-1)^F Q^2 e^{-\b H}  -   {\partial\over \partial x^i} \tr \, e_i
(-1)^F Q e^{-\b H}, \cr }$$
we find that the $\b$ variation vanishes up to a total divergence. In this
expression, $ e_i {\partial\over \partial x^i} $, is the derivative term in the
supercharge, $Q$. If we define $e_n$ to be the fermion in the normal direction
to the boundary, then the index can be written as a sum of two terms,
\eqn\twoterms{ {\rm Ind} = \lim_{R\r\infty} \lim_{\b_o\r 0} \{ \int_{|x|<R}{
\tr \, (-1)^F e^{-\b_o H}} + {1\over 2} \int_{|x|=R} \int_{\b_o}^{\infty}{ d\b
\, \tr \,
e_n  (-1)^F Q e^{-\b H}} \}.  }
At first sight, keeping track of these various limits may seem like a
technicality; however, that is not the case. So, rather than continue a general
discussion, let us revisit an old friend to see how these manipulations work
concretely. A number of the diffculties that arise in the D-particle cases will
become clearer.

\subsec{The harmonic oscillator revisited}

Let us consider a single supersymmetric harmonic oscillator, which has a unique
ground state, and a discrete spectrum. The supercharge is given by,
$$ Q = \p_1 p + \p_2 x, $$
where $ \p_1^2 = \p_2^2=1$ and $ \{ \p_1, \p_2 \} =0$. The Hamiltonian is
one-half the square of the supercharge,
$$ \eqalign{ H &= {1\over 2} Q^2 \cr
&= {1\over 2} (p^2 + x^2 - i \p_1 \p_2).} $$
Now to evaluate the principal term, we can consider the first term in
\twoterms, we which can write as,
$$ \int_{-R}^{R}{ dx {1\over \sqrt{2\pi \b_o}} \, \tr \, i\p_1\p_2 e^{-
{\b_o\over 2} ( x^2 - i \p_1\p_2)} + \ldots,}
$$
where $(-1)^F = i\p_1\p_2$ in this case, and squares to the identity. The
omitted terms are suppressed by powers of $\b$. Evaluating the trace on the
fermions, or in the equivalent path-integral language, integrating out the
fermion zero modes, gives a leading contribution in $\b_o$,
$$  \int_{-R}^{R}{ dx {1\over \sqrt{2\pi \b_o}} \b_o e^{-\b_o x^2/2},} $$
which gives,
$$ \int_{ - R \sqrt{\b_o/2}}^{R \sqrt{\b_o/2}}{ dx {1\over \sqrt{\pi}}
e^{-x^2}.}$$
If we take $\b_o$ to zero faster than $R^{-2}$, this term vanishes, while if we
take $\b_o$ to zero more slowly, we obtain the expected answer of one. Whether
or not we get a contribution from this term depends on how we choose to take
$\b_o$ to zero. When this term does not contribute, the second term in
\twoterms\ contributes, and the principal contribution is shifted to a boundary
term as we shall see. In this model, the principal contribution is the only
contribution to the index.

The boundary term gets two equal contributions from $R$ and $-R$ in this case,
and so can be written,
$$ {1\over 2}\int_{\b_o}^\infty{d\b \, \tr \, (-i \p_1)(i \p_1\p_2) Q e^{-\b H}
}
\Bigr|_{x=R}. $$
As $R$ becomes large, the potential term $e^{-\b V}$ damps the kernel, $e^{-\b
H}$, for large $\b$. We therefore do not need non-perturbative information
about the kernel to evaluate this contribution -- a small $\b$ approximation
suffices. Whenever there is a mass gap, we have this nice damping, which is the
reason that the index is usually computable.

In this case, evaluating the trace on fermions gives,
$$  \int_{\b_o}^\infty{d\b \, x {1\over \sqrt{2\pi \b}} e^{-\b x^2/2} }
\Bigr|_{x=R},  $$
and on rescaling, we obtain:
$$ \int_{\b_o R^2/2}^\infty{d\b \, {1\over \sqrt{\pi \b}} e^{-\b }. } $$
Now we see that this term can contribute if $\b_o$ is taken to zero
sufficiently quickly with $R$. Of course, the $L^2$ index is one regardless of
how fast or slowly we choose to take $\b_o$ to zero. In the case of a model
with potential $V$ homogeneous in $x$ with degree $k$, a similar argument
can be applied. In that case, if we take $\b_o$ to zero slower than $R^{-k}$
then the
principal contribution is localized to the first term of \twoterms, while if
take $\b_o$ to zero faster than $R^{-k}$, the second term contributes. In the
following section, we shall evaluate the  principal contribution for the two
D-particle case by letting $\b_o$ go to zero more slowly than $R^{-4}$. This
seems computationally simpler than trying to localize the contribution to the
boundary.

\subsec{Reducing the principal term to quadrature}

To evaluate the principal contribution, we have to construct a reasonable
approximation to $
e^{-\b H}$. We will not need to alter the usual perturbative construction of
the partition function because of the flat points of the potential, $ V$. We
start by writing,
$$ e^{-\b H} = {1\over 2\pi i} \int_{\g}{ e^{-\b z} {1\over H-z} dz,} $$
where $ \g$ is a contour enclosing the spectrum of H. Let us consider
the generic
situation away from the flat points. We can approximate $ (H-z)^{-1}$ by a
perturbation
series,
\eqn\approx{ {1\over H-z} (x,y) = \int{ { e^{i k \cdot (x-y)}\over (k^2/2 +
V - z)} (1 - {H_F
\over (k^2/2 + V - z)} + \ldots)},}
where the first correction, proportional to $ H_F$, is shown, and subsequent
terms are
constructed iteratively in powers of $(k^2/2 + V - z)^{-1}$.  The
corresponding propagator
takes the form,
\eqn\heat{ e^{-\b H} (x,y) = {1 \over (2\pi \b)^{l/2} } e^{-{|x-y|^2 \over 2
\b}}e^{-\b V}
e^{-\b H_F} + \dots ,}
where $ l = d (N^2-1)$ is the dimension of the space of scalars.  This
approximation is
reasonable for small $ \b$. The omitted terms which correct this
approximation appear with a
higher power of $(k^2/2 + V - z)^{-1}$, in \approx, and consequently give
rise to terms
suppressed by powers of $ \b$ in \heat. This approximation then suffices for
evaluating the
first term in \twoterms, where we choose the take $\b$ to zero more slowly than
$R^{-4}$. Substituting the leading approximation for the
propagator gives,
$$
\eqalign{
 \phantom{=}& \lim_{R\r\infty}\lim_{\b \rightarrow 0} \int_{|x|<R}{
\int_{SU(N)} dt  \, \tr \,
(-1)^F e^{-\b H} \Pi (g) \, (gx,x) } \cr
=& \lim_{R\r\infty}\lim_{\b \rightarrow 0} \int_{|x|<R}{\int_{SU(N)} dt \, \tr
\,
(-1)^F {1 \over (2\pi \b)^{l/2} } e^{ - | i \vec{t} \cdot\vec{C}^b x|^2
\over 2 \b} e^{-\b V}
e^{-\b H_F} e^{it_A C^f_A} + \dots,}\cr
} $$
where we have approximated the group element by $ I + i
\vec{t} \cdot\vec{C}^b$, for reasons explained in the previous section. This
amounts
to localizing the integral over the gauge group to a small neighborhood of
the identity -- a slight twist on the usual localization to the minima of
the potential.  As usual, the inclusion of $ (-1)^F$ forces us to absorb
fermion zero modes. In our trace, $
(-1)^F$ is realized as the volume form for the Clifford algebra \fermions.
There are two sources for fermions: the first is from  $ e^{-\b H_F}$, while
the second source is the fermion projection operator, $ e^{it_A C^f_A}$,
inserted into the trace. After writing the fermion term, $ it_A C^f_A - \b
H_F$, as $\psi M \psi $ for some matrix $M$,
the trace over the Clifford factors gives the Pfaffian of $ M$, which is a
polynomial in $ x$ and $ t$. On rescaling the integral, we obtain,

\eqn\finalint{   \int dt \,  \int dx \,  {1\over ( 2\pi )^{l/2}} e^{- | i
\vec{t}
\cdot\vec{C}^b x|^2 / 2} e^{ - V(x)}  \, {\rm Pf}(M), }
where the integration region for  $ t$ is now $ {\bf R}^{N^2-1}$, while
for $ x$, the region
is $ {\bf R}^{d (N^2-1)}$. As will be clear from the subsequent explicit
computation, the Pfaffian is of definite sign when $d$ is odd, and the
integral is thus non-vanishing. However, it is far from clear that this term
yields an integer, and indeed, it generally is not integral. Therefore,
there had better be a non-vanishing correction term.

We stress again that it is very natural to consider $ t$ on equal footing
with the coordinates $ x^i$.
Let us denote $ t$ by $ x^0$, and define $ \g^0 $ to be $ i  I$, where $
I$ is the identity
matrix. The coordinates $ x^i_A$ now form an $ (N^2-1) \times (d+1)$ matrix.
In this notation, the matrix takes the form $M = -(i/2) f_{ABC} x^i_B \g^i$,
and the integral admits an $SO(d+1)$ symmetry which we shall use in section
four to compute explicit values for this term in the
two-particle case.

\subsec{The non-Fredholm case}

When the Hamiltonian under consideration has continuous spectrum, the twisted
partition function is generally $\b$-dependent. The heuristic reason for the
$\b$-dependence is that the density of states for the bosonic and fermionic
scattering states can differ. Supersymmetry pairs bosonic and fermionic modes,
but does not necessarily preserve the spectral density. In these cases, the
principal contribution to the index is not necessarily integer, and there must
be an additional contribution from the second term in \twoterms. In the case
where there is a mass gap, this contribution can be perturbatively evaluated.
What happens in the case where there is no mass gap?

Let us choose a real supercharge, $Q$, which squares to the Hamiltonian up to a
gauge transformation. For this discussion, we will set the gauge constraints to
zero. $Q$ is then a self-adjoint elliptic first-order operator, which
anti-commutes with our ${\bf Z}_2$ involution, $(-1)^F$. Let us define $Q_+$ as
the restriction of $Q$ to the $+1$ eigenspace of $(-1)^F$, i.e. the bosonic
states. It may be helpful to think of $Q$ as a matrix,

$$ \left (\matrix{
0 & Q_+^*  \cr
Q_+ & 0  \cr
}\right ) , $$
where $Q_+^* = Q_-$ is the restriction of $Q$ to fermionic states. The
computation that we need to perform is the calculation of the $L^2$ index of
$Q_+$. This operator, though elliptic, is not Fredholm. Recall that
a Fredholm operator, by definition, has a finite-dimensional kernel and
cokernel. The fact
that the continuous spectrum of $H=Q^2$ contains scattering states with
arbitrarily small energies implies that the image of $Q_+$ is not closed, and
the cokernel is infinite-dimensional, and distinct from the
 kernel of $Q_-$. Hence, we take the $L^2$ index of $Q_+$ to be the dimension
 of
$\ker (Q_+)\cap L^2$ minus the dimension of
$\ker (Q_+^*)\cap L^2,$ which is not, in this case,
 the dimension of the kernel minus the dimension of the cokernel of $Q_+$.

Let us consider how to compute this index. Suppose that there exists a Green's
function, $G$, for
$Q^2$; i.e. a self-adjoint singular integral operator
$G$ which annihilates the kernel
of $Q^2$ and acts as $(Q^2)^{-1}$ on the orthogonal complement of the kernel.
By definition, $G$ obeys,
$$Q^2G = I - P,$$
where $P$ denotes the orthogonal projection onto the kernel of $Q^2.$ So, $P$
then annihilates all states which are not zero-energy.
We recall that by singular integral operator we mean an operator  which
is obtained by integrating against a matrix-valued kernel,
$g(x,y)$, which has a well-understood singularity along the diagonal
$x=y$. For example, the inverse of the Laplacian in three dimensions is the
familiar kernel, $\sim |x-y|^{-1}$.
Let us denote the restriction of $G$ to the $+1$
eigenspace of $(-1)^F$ by $G_+$, where:
$$G_+ = G (I+(-1)^F)/2. $$
Let $P_{\pm}$ denote the orthogonal projection onto
the $L^2$ kernel of $Q_{\pm}$.
Then $QQG_+  = (I-P)(I+(-1)^F)/2 = (I+(-1)^F)/2 - P_+ ,$ and
similarly, using  the fact that $Q$ anti-commutes with $(-1)^F$ and
commutes with $G$, we have
  $QG_+Q  = (I-(-1)^F)/2 - P_- .$
 Therefore, the index of $Q_+$ can be expressed as,
$$ \eqalign{ {\rm Ind} Q_+ &= \tr P_+ - \tr P_- \cr &=\tr [(I+(-1)^F)/2 -
QQG_+] - \tr[(I-(-1)^F)/2 - QG_+Q] \cr &=\tr((-1)^F - [Q,QG_+]) \cr & = \tr(
(-1)^F + [Q,(-1)^F QG/2]).}$$

Of course, the difficulty is that we cannot construct $G$ explicitly --
even the claim that $G$ is represented by a singular integral operator
requires some justification, which we will give shortly. This difficulty can be
summarized in the following way: given a set of eigenstates, $\p_{E\a}$, with
eigenvalue $E$ under $Q^2$, the inverse is formally,
$$ (Q^2)^{-1} = \sum_{E\a,E>0} E^{-1} \p_{E\a}(x) \p_{E\a}(y), $$
where the sum  may be over continuous indices. When the
continuous spectrum is not bounded away from zero, it is not clear that
the resulting sum converges to a function in any reasonable sense, since $G$ is
 an unbounded operator on $L^2$ wavefunctions in this case. However, if the
scattering states do not pile up at low-energies, then it should be intuitively
reasonable that $G$ is still nice, as, for example, in the free-particle case.
We will see this later by realizing
$G$ as a limit of bounded singular integral operators $G_w$. Physically, this
limiting procedure is equivalent to adding a mass term to the propagator, and
taking the limit where the mass vanishes. The problem we are decribing is a
common one in any theory with massless particles.
 So, let us proceed
along the usual path by
constructing an explicit approximation $W$ to $G$ for which we can compute
the trace,
$$\tr( (-1)^F + [Q,(-1)^F QW/2]).$$
We must then verify that for a carefully constructed $W$ this trace
is the same as the one computing the index of $Q_+$.

 Our approximation will have the property that,
$$Q^2W = I-E, $$
for some compact error term $E$ given by integrating against a matrix-valued
kernel, $e(x,y)$. We also want $e(x,y)$ to decay polynomially
in $(|x|+|y|)$ to a sufficiently high power which we need to determine, and
 study in greater detail later.

Let us describe what data is needed in order to insure that
$\tr( (-1)^F + [Q,(-1)^F QW/2])$ computes the index. This is equivalent to the
vanishing of  $\tr[Q,(-1)^F Q(G-W)/2]$, which is  the difference between
 $\tr ((-1)^F + [Q,(-1)^F QG/2])$, which computes the index, and $\tr((-1)^F +
[Q,(-1)^F QW/2])$, which we hope computes the index.

 For any integer $m$, the operator $W$ can be constructed so that
the  kernel for $G-W$ has $m$ continuous derivatives.
Let $\chi_R$ be the characteristic function of a ball $B_R$ of
radius $R$. The characteristic function is defined to be one on the ball and
zero
elsewhere. We will throw $\chi_R$ into the various traces to serve as a cut-off
on the infra-red physics.
 By a similar argument to the one used in the section 3.1, we may
use the divergence theorem to transform
$\tr \chi_R [Q,(-1)^F Q(G-W)/2]$ into an integral over the boundary of $B_R$.
Therefore, if we can show that $(-1)^F Q(G-W)(x,x)$ is decaying sufficiently
rapidly
at large $|x|=R$, we deduce that $\tr\chi_R[Q,(-1)^F Q(G-W)/2]$ converges to
zero,
and the index can be computed by replacing $G$ by our approximation, $W$.

 To prove that $Q(G-W)$ decays sufficiently quickly, we will need to use an
argument that may be unfamiliar to the reader which establishes a
correspondence between asymptotic estimates for $Q^2$ and decay rates for
solutions $\psi$ to $Q^2\psi = F,$ where $F$ satisfies some growth constraint.
In order to orient the reader, let us first examine what the argument says in a
much simpler case.
Consider the differential equation in one variable $r$, on $(0,\infty),$

$$(-\frac{d^2}{dr^2} + w^2)f = g,$$
where $w$ is some constant. For this equation, a weak form of our general
argument below says that if
$e^{awr}g\in L^2,$ for some $a\in (-1,1),$ then we may conclude that $e^{awr}
f$ is normalizable, if $f$ satisfies the growth
 constraint
$e^{-bwr}f\in L^2,$ for some $b<1.$ The condition on the growth of $f$ is
clearly necessary to rule out
the addition of the non-normalizable $e^{rw}$ to any solution.  In this simple
case, we
can prove this result by a direct integration. We will use the following
analogous result, which is established in much greater generality in \rAg.

    Suppose that for some positive constant $c$ and some compact set $K$,
we have an estimate of the following form:
$$\|Qf\|^2 \geq \|cf/r\|^2,$$
for all wavefunctions, $f$, which vanish outside $K$. With these assumptions,
 if $Q^2F = e,$ with $r^{c}e\in L^2$, and if $F$ satisfies the growth
constraint that $F / r^{c-s} $ be normalizable for some positive $s$, then
$${r^{c-1}\over\ln(r)^{1+\epsilon}} F\in L^2,$$
 for all positive $\epsilon$. Also,
$${r^c\over \ln(r)^{\epsilon}} QF\in L^2,$$
 for all positive $\epsilon$.
In the familiar case where we have $c$ instead of $c/r$, we would obtain
exponential decay as in the one-dimensional example. The
reason we have a weaker decay rate in this example is that the decay
is roughly no worse than $e^{-\psi}$, where $\psi$ is a function with
$|d\psi|^2$ less than the asymptotic lower bound for $Q^2$, which is $c^2/r^2$
in this case, and $w^2$ in the one-dimensional example.

For purposes of illustration, let us sketch a proof of these growth
estimates. For any function $u$ supported in the complement of K, integration
by
parts yields
$$0 = (Q^2F,u^2F) = \|QuF\|^2 - \|[Q,u]F\|^2.$$
The assumed estimate implies that,

$$\|cuF/r\|^2 \geq \|[Q,u]F\|^2.$$
{}From this formal inequality, we can deduce the normalizability of
$cuF/r$ when $cu/r > |[Q,u]|.$  More accurately, we can deduce
the integrability of $(c^2u^2/r^2 - |[Q,u]|^2)F^2.$ Let us define a cutoff
function $\rho$ which is identically one outside a large
ball $K'$ containing $K$ in its interior, and vanishing in $K$.  Then taking
$u$ to be $r^c/\ln(r)^{\epsilon}$
times the cutoff function $\rho$ gives formally,
$$\|cr^{c-1}F/\ln(r)^{\epsilon}\rho\|^2 \leq \|(r^{c-1}(c-\epsilon/\ln(r))\rho
+ r^c\rho')/\ln(r)^{\epsilon}F\|^2.$$
Collecting terms, we obtain:
$$\int r^{c-2}(2c\epsilon-\epsilon^2/\ln(r))|F|^2/\ln(r)^{1+\epsilon}\rho\|^2
 \leq
c_{K'}\int_{K'}|F|^2,$$
for some constant $c_{K'}$, which depends on $K'$ and $c$.
A limiting argument, approximating $r$ by a sequence of bounded functions
 can be used to obtain from this formal inequality the boundedness of the left
side, when $|F|$ is bounded on compact sets.

We will need the following variant of this inequality.
Suppose that
$$\tilde{Q}^2 = -{\partial^2\over \partial r^2} - {(2d-1)\over
r}{\partial\over\partial r} +  w,$$
where
 $w$ is now a positive operator with $w\geq c^2/r^2.$
Suppose that $\tilde{Q}^2F = 0$ in the complement of a compact set $K$.
For some $k$ to be determined, consider the point where $r^kF$ is maximum.
By the maximum principle,
we have at the maximium that, $F' = -kF/r$ and
$  (r^kF)'' \leq 0$, where:

$$ \eqalign{ (r^kF)'' = & k(k-1)r^{k-2}F + 2kr^{k-1}F'+r^kF'' \cr
 \geq & k(k-1)r^{k-2}F - 2k^2r^{k-2}F-(2d-1)r^{k-1}F'+r^{k-2}c^2F  \cr
 =  & (2d-k-2 + c^2/k)kr^{k-2}F.} $$
If we choose $0<k< 2(d-1),$ we may deduce that if the maximum exists it must
occur
in $K$.  If $Fr^{a}$ is bounded for any positive $a$, we may deduce that
$Fr^k$ is in fact bounded by its values on $K$ for all $k$ with
$k< 2d+c^2/k-2$.  In our applications, $c^2$ will usually be the
first or second eigenvalue of the standard spherical Laplacian, which is
$0$ or $(d-1).$ We recall that the standard
Laplace operator on $S^{d-1}$ has eigenvalues, $k(d+k-2)$, with $k =
0,1,2,\ldots$, where the multiplicity of each eigenvalue is,
${(2k+d-2)(k+d-3)!\over{k!(d-k)!}}.$
 In the first case, we see that we get $r^{d-2}$ decay; in the second we see
that for $d$ at least $3$,
we get faster than $r^{2-2d}$ decay. These estimates extend easily to the
case when $\tilde{Q}^2F = e,$ where $e$ satisfies the condition that $r^{2d}e$
is bounded.
We will apply this to the case when $\tilde{Q}^2 = Q_x^2 + Q_y^2,$
the hamiltonian in the first and second variables and
$F$ will be a kernel constructed the difference between the Green's function
and $W$. This
argument formalizes the observation that the product of two
elements of the kernel of $H$ should decay twice as fast as
a single element of the kernel, and $G-W$ looks like such a  product.

We will show that,
$$Q^2 = \Delta_r + u/r^2,$$
acting on wavefunctions supported outside a large compact set; here $r$ is the
distance along the flat direction, $\Delta_r$ is the radial part of the
Euclidean Laplacian
in the flat directions, and the operator $u$ is semi-positive with first
eigenvalue greater than the first or second eigenvalue of the
spherical Laplacian. We can then apply the above argument to deduce that
$G-W$ decays like $r^{2-2d}$.  We can improve this estimate by observing
that we get the second eigenvalue when the operator is
restricted to wavefunctions with odd parity, and the first eigenvalue for the
restriction to even parity wavefunctions.
Split $(G-W)$ into its even and odd components.
The odd component of $(G-W)$ decays faster than $r^{1-d}$ by the preceding
discussion, and it is easy to see that applying $Q$ only improves the decay
rate.  To see this, use the inequality for $Q^2F= 0$,
$$\|wQF\|\leq 2\|[Q,w]F\|,$$
and choose $w$ appropriately.
For the even component of $G-W$, we have that its image under $Q$ is odd;
therefore,
we again have the improved decay rate determined by the second eigenvalue
of the sphere.
This estimate requires that
$(G-W)r^a$ be bounded for some positive $a$ and that $G$ be given as
a singular integral operator. The last condition is needed to ensure that
$(G-W)$ is a smooth function.
We will not prove these results in detail but will merely sketch
how they follow from the same sequence of ideas we have introduced.
One considers first instead of $G$ and $W$ corresponding
operators $G_w$ and $W_w$, where $G_w$ is $(Q^2+w)^{-1}$
restricted to the orthogonal complement of the kernel of $Q^2$ etc.
It is easy to get the desired initial $r^a$ boundedness for
$G_w - Q_w$ for each $w>0$. One
then can get the desired growth bound; i.e. we show that the max is
controlled by
an estimate on the compact set. We then allow $w$ to tend to zero to get
the desired result for $G-W = \lim_{w\rightarrow 0} (G_w-W_w).$
  This type of argument can also be used to show that $G$ is a
singular integral operator.

This then establishes the
desired decay estimate given one: a construction of $W$ which leads to
sufficiently small $E$; that is $|e(x,y)|\leq (|x|+|y|)^{-d-1},$
 and two: a demonstration of the claimed asymptotic lower bound for $Q^2$.
Under these two conditions, $\tr((-1)^F + [Q,(-1)^FQW/2])$ computes the index.
We now turn to the evaluation of $[Q,(-1)^FQW/2]$, which we need to boil the
problem down to a concrete computation.

It will be convenient to arrange the construction of $W$ so that on a large
compact set, say a ball $B_R$, its contribution to the index can be computed by
the standard
principal term computation described in the previous subsection. On the
complement of this set we will need to use
special coordinates to find a nice expression for $W$. This
requires us to define an approximation $A$ to $QG$ rather than
the approximation $W$ to $G$, but since, off a compact set, $A$ will clearly
be of the form $QW$, this will not affect our preceeding discussion.
The use of  two separate constructions for $W$ in different
regions may seem, perhaps, a bit unnatural when dealing with Euclidean space.
 It is forced on us, in part, by
the need to obtain very good control of the error term $E$ as
$|x|$ tends to $\infty$. This rules out the use of the local
computation used in $B_R$ and described previously. Moreover, the
 special coordinates we use in the complement of the compact set, like
polar coordinates, become singular at the origin. Therefore, we will need to
use
two sets of cutoff functions to patch the two approximate inverses together.

Let $\rho_{n,j}(x)$ be a sequence of cutoff functions which approach
$\chi_{jR}(x)$,
and set $\rho_n = \rho_{n,1}.$
Let $W'$ be our approximate Green's function near $\infty$, which we will
construct in section five. The operator
$\int_0^{\b_0}d\b \, Qe^{-\b Q^2}(x,y)$ is the standard kernel
that we will use in $B_R$.
We can create a global approximation to $QG$ by defining:
$$A(x,y) = \rho_{n}(x)\int_0^{\b_0}d\b \, Qe^{-\b Q^2}(x,y) \rho_{n,2}(y) +
(1-\rho_{n}(x))QW'(x,y)(1-\rho_{n,1/2}(y)).$$
The cutoff functions on the left of each operator are inserted to
average the two operators. The cutoff functions on the right, however,
are inserted so that the operators are localized to the domains where
they are well-defined, and satisfy the desired estimates.  The right cutoffs
are of course chosen to be identically one on the support of the left
cutoffs; otherwise, they would destroy the averaging effected by the
left cutoffs. This is the reason for the second index on $\rho$.

 Then to evaluate $[(-1)^FA, Q]$, we write:
$$ \eqalign{ [(-1)^F A, Q] =& [Q,\rho_{n}](-1)^F \int_0^{\b_0}d\b \, Qe^{-\b
Q^2}\rho_{n,2}  - [Q,\rho_{n}](-1)^F QW'(1-\rho_{n,1/2}) \cr &
- \rho_{n}(-1)^F(I-e^{-\b_0 Q^2})\rho_{n,2} +
(1-\rho_{n})(-1)^F(I-E)(1-\rho_{n,1/2}) \cr &
+ (1-\rho_{n})(-1)^F Q[Q,W'](1-\rho_{n,1/2}) + \ldots, \cr
= & [Q,\rho_{n}]\int_0^{\b_0}d\b \, Qe^{-\b{ Q^2}}  - [Q,\rho_{n}](-1)^F QW'
\cr &
- (-1)^F (I-\rho_{n}e^{-\b_0 Q^2}) + (1-\rho_{n})(-1)^F(-E) \cr &
+ (1-\rho_{n})(-1)^F Q[Q,W'] +  \ldots, }$$
where the omitted terms are terms that trace to zero.
We will construct $W'$ in section five so that the trace of
$(1-\rho_{n})(-1)^F(-E) + (1-\rho_{n})(-1)^F Q[Q,W']$ will tend to zero
as $n $ tends to $\infty$.  Thus subtracting off
the $(-1)^F I$ term we are left to compute,
$$ \tr[Q,\rho_{n}]\int_0^{\b_0}d\b\, Qe^{-\b Q^2}  - [Q,\rho_{n}](-1)^F QW'
+ (-1)^F \rho_ne^{-\b_0 Q^2}.$$
The last term is the principal term which is given by evaluating the integral
\finalint.
Taking the limit as $n$ tends to $\infty$, the
two commutator terms converge to the boundary traces,

$$ \int_{|x|=R} \left( \tr\, e_n\int_0^{\b_0}d\b\, Qe^{-\b Q^2}  - \tr\,
e_n(-1)^F QW' \right) .$$
Choosing $\b_0$ to go to zero more slowly than $R^{-4}$, the integral,

$$\int_{|x|=R} \tr\, e_n\int_0^{\b_0}d\b\, Qe^{-\b Q^2}, $$
decomposes into two pieces. One is associated with a small neighborhood of the
flat regions, which consist, say, of all points of distance at most
one from a flat point. The other contribution is associated with the
complementary region, i.e. almost all of the sphere of radius $R$. It is not
difficult
to show that the contribution from the flat region is squeezed to zero
as $R$ tends to $\infty$. The
contribution from the complementary region is not vanishing. Using similar
arguments to those presented earlier in this section, it is possible to show
without extensive computation that
this term exactly cancels the principal term. Standard constructions
give a $W'$ whose  contribution
from the complementary region of
$\tr\, e_n(-1)^F QW'$ also exactly cancels the contribution
of $\int_{|x|=R} \tr\, e_n\int_0^{\b_0}d\b \,Qe^{-\b Q^2} $. Therefore, the
total
contribution of this boundary integral to the index comes from,
\eqn\boundary{ -{1\over 2}\int_{N_F(R)}\tr\, e_n(-1)^F QW',}
where $N_F(R)$ is a small neighborhood of the flat points on the boundary of
the space, which is a sphere
of radius $R$. We will show in section five that this integral converges to
$-1/4$ in the two-particle case. In summary, the additional contribution to the
index from the boundary is computed by evaluating \boundary, which is localized
to the flat directions. The sum of \boundary\ and the usual principal term from
\finalint\  must then be integer.

\newsec{Two-Particle Binding}

\subsec{Symmetries and Pfaffians}

The simplest case to consider is $ N=2$; already, however, the integral
\finalint\ is thirty-dimensional for type IIA zero-branes!  We shall have to
use the various
symmetries available to us to simplify the computation of the principal term.
Recall that the
coordinates $ x^i_A$ now form a $ 3 \times (d+1)$ matrix. The integral
\finalint\ is
invariant under the symmetry, $ x \rightarrow gxh$, where $ g$ is an
element of $ SO(3)$
acting on the left while $ h$ is an element of $ SO(d+1)$ acting from the
right on $ x$.  Note that the left action is a gauge transformation.

By using these symmetries, we shall rotate $ x$ into a special form,

\eqn\coord{ \left (\matrix{
b_1 & 0 & 0 \cdots & & & 0 \cr
0 & b_2 & 0 & \cdots & & 0 \cr
0 & 0 & b_3 & 0 & \cdots & 0 \cr
}\right ) , }
and reduce our integral to one over only three variables.

In these special coordinates, the potential (including the gauge parameters)
takes the special form,

$$   {\widetilde V}  = - {1\over 2}  (  b_1^2 b_2^2+ b_1^2 b_3^2 + b_2^2
b_3^2 ). $$
We shall now evaluate $ \Pf (M)$ by evaluating the determinant of $ M$.
For the moment, let us
return to general coordinates where $ M = -(i/2) f_{ABC} x^i_B \g^i $. For
convenience, let us
denote $ x^i_A \g^i$ by $ x_A$. The matrix then takes the form,

$$
M = {i \over 2 } \pmatrix{
0 & x_3 & -x_2 \cr
-x_3 & 0 & x_1 \cr
x_2 & -x_1 & 0 \cr
}.
$$
By row manipulations, or equivalently, by studying the eigenvalue
equation, we find that the
determinant can be expressed as,

$$
\det (M) = {1\over 2^{3(d-1)}} \det (x_1 x_2 x_3) \det (1- x_1 x_2^{-1} x_3
x_1^{-1} x_2
x_3^{-1}),
$$
where $ x_A^{-1} = ( x_A - 2i x^0_A  I ) / | x_A |^2 $. After rotating $
x$ into our
convenient set of coordinates \coord, we can compute this determinant, and
on taking the square root obtain,
$$ \Pf (M) = {1\over 2^{2(d-1)}} (b_1 b_2 b_3)^{d-1} .$$
We can immediately see that when $d$ is odd, the
Pfaffian is an even function of the variables in both special and general
coordinates, and the corresponding integral \finalint\ is non-vanishing.

The last ingredient that we require to compute the integral \finalint\ is
the measure for our simplified coordinates \coord.  We shall obtain the
measure by gauge-fixing the integral \finalint\ using
the Faddeev-Popov approach. Let us take $ x'$ to  be,

$$
\left (\matrix{
b_1' & 0 & 0 \cdots & & & 0 \cr
0 & b_2' & 0 & \cdots & & 0 \cr
0 & 0 & b_3' & 0 & \cdots & 0 \cr
}\right ) .
$$
We shall insert one into the integral \finalint\ in the form,

$$ \int db'  dg dh \, \delta ( x' - gxh) f(b), $$
where we have to determine $ f(b)$. For some $ g_0$ and $ h_0$, $ x_0 =
g_0 x h_0$ takes the
form \coord. The integrals over $ g$ and $ h$ then reduce to integrals in
a small neighborhood
of $ g_o$ and $ h_o$, with the exception of the $ SO(d-2)$ subgroup of $
SO(d+1)$ that leaves the
form \coord\ invariant. If $ T$ is a generator for the left $ SO(3)$
action, and $ R$ a
generator of the right $ SO(d+1)$ action which does not leave \coord\
invariant, then we can
replace integration over $ g,h$ by,
$$  \eta(d) \,  \vol (SO(d-2)) \int db'  dT dR \, \delta ( x' - x_o - T x_o
- R x_o) f(b).  $$
\noindent
The remaining integrals are straightforward, and we find that,

$$ f(b) =
{1\over \eta(d) \, \vol
(SO(d-2))}(b_1b_2b_3)^{d-2} | (b_1^2 - b_2^2)(b_1^2-b_3^2)(b_2^2-b_3^2)|.$$
The integrals over $b$ are constrained such that $ b_1>b_2>b_3$. The
symmetry factor $\eta(d)$ is 4 for $d>2$, but is 2 for $d=2$ because the
left and right symmetry groups are then both $SO(3)$. The value of the symmetry
factor can also be checked by computing a Gaussian integral in the
$d$-dimensional model, and comparing the result to the answer obtained using
the
measure for these special coordinates.

Finally, inserting one
in this form into the integral \finalint, and
integrating over $ x,g,h$
gives for $d>2$,
$$  {1 \over \vol (SO(3))} {1\over (2 \pi)^{3d/2}}\int db \, {\vol(SO(d+1))
\vol(SO(3)) \over 4\, \vol(SO(d-2))} (b_1b_2b_3)^{d-2} | (b_1^2 -
b_2^2)(b_1^2-b_3^2)\times $$ $$ (b_2^2-b_3^2)| {1\over 2^{2(d-1)}} (b_1 b_2
b_3)^{d-1} e^{ {\widetilde V}}. $$
The first factor of $1/\vol(SO(3))$ comes from the normalization of the
integration over the gauge group, where we recall that we chose the
normalization so that $ \int_{SU(2)} dt =1 $ prior to rescaling.

\subsec{Computing the principal contribution}

Now there is a nice change of variables that will allow us to evaluate this
integral. Set: $y_1= b_2b_3/\sqrt{2}, y_2=b_1b_3/\sqrt{2},
y_3=b_1b_3/\sqrt{2}$, and the integral becomes,
$$ \eqalign{ \phantom{=} & \int db \, (b_1b_2b_3)^{d-2} | (b_1^2 -
b_2^2)(b_1^2-b_3^2)(b_2^2-b_3^2)|  {1\over (2 \pi)^{3d/2}} {1\over
2^{2(d-1)}} (b_1 b_2 b_3)^{d-1} e^{ {\widetilde V}} \cr
=& \int dy (y_1y_2y_3)^{d-3}| (y_1^2 - y_2^2)(y_1^2-y_3^2)(y_2^2-y_3^2)|
{1\over 2}{1\over 2^{2(d-1)}}{1\over \pi^{3d/2}}e^{-y_1^2+y_2^2+y_3^2} \cr
=& \int_{{\bf R}^{3d} } dx e^{-|x|^2}{1\over 2}{1\over 2^{2(d-1)}}{1\over
\pi^{3d/2}} { \eta(d-1)\,\vol(SO(d-3))\over \vol(SO(d)) \vol(SO(3))} \cr
=& {1\over 2}{1\over 2^{2(d-1)}}  {\eta(d-1)\,\vol(SO(d-3))\over
\vol(SO(d)) \vol(SO(3))}.
}  $$
Lastly, we must multiply the result by the value of $\tr(I)$ from the trace
over the fermions, which gives an extra factor $2^{3(d-1)}$. The net result
is the formula:

\eqn\bulk{ P = 2^{d-2} { \eta(d-1) \, \vol(SO(d+1)) \vol(SO(d-3)) \over
\eta(d) \, \vol(SO(d-2))\vol(SO(d))\vol(SO(3))},}
for the principal contribution, $P$, for $d$ odd, where we recall that $\vol
(SO(n)) = \vol
(S^{n-1})
\cdots \vol (S^1)$, and that $\vol (S^n) = 2 \pi^{n+1\over 2}/\Gamma({n+1\over
2})$. Let us conclude
this discussion by listing the explicit values in the following table:
\vskip 0.2in
{\vbox{\ninepoint{
$$
\vbox{\offinterlineskip\tabskip=0pt\halign{\strut\vrule#
&\hfil~$#$~\hfil &\hfil~$#$~\hfil &\vrule#
&\hfil ~$#$~ \hfil&\hfil ~$#$~ \hfil&\hfil ~$#$~ \hfil&\vrule#
&\hfil ~$#$~ \hfil&\hfil ~$#$~ \hfil&\hfil ~$#$~ \hfil&\vrule#
&\hfil ~$#$~ \hfil&\hfil ~$#$~ \hfil&\hfil ~$#$~ \hfil&\hfil ~$#$~ \hfil
&\vrule#\cr\noalign{\hrule}
&  {\rm Dimension}  &&  & {\rm Principal \,\, Contribution}  & &&\cr
\noalign{\hrule}
\noalign{\hrule}& 3 && & 1/4 & &&\cr
\noalign{\hrule}& 5 && & 1/4 & &&\cr
\noalign{\hrule}& 9 && & 5/4 & &&\cr
\noalign{\hrule} }
}$$ {\cl {{\bf Table I}: The principal contribution to the index.}}}
\vskip7pt}}

\newsec{The Propagator for Well-Separated Branes}

\subsec{Some general comments}

We have determined in the previous section that the principal contribution to
the index is fractional. Since the index must be integer, there is a missing
contribution. The manner in which this contribution arises is quite surprising,
and involves a bizarre conspiracy of cancellations. Let us outline the
procedure we will follow before presenting a detailed discussion. We will
construct an approximation to the propagator for the two zero-branes when they
are far apart. The approximation will be sufficiently good in the sense that
any corrections will not contribute to the boundary term \boundary. At long
distances, the only states that make a sizable contribution to the propagator
are those localized along the flat directions of the potential. The simplest
approximate description of the physics governing the light degrees of freedom
is in terms of free particle propagation along the flat directions. This is
immediately modified when we try to `integrate' out the massive modes. Let us
label coordinates for the $d_a$ massive directions by $y$. Then for example,
the action of $\partial_r^2$, which is part of the Laplacian \free, on the
wavefunction for the flat direction is modified because of its action on the
harmonic oscillator ground state,

$$ \eqalign{ \partial_r^2 \left( {r\over \pi} \right)^{d_a/4} e^{-r |y|^2/2} =&
\left( {r\over \pi} \right)^{d_a/4} e^{-r |y|^2/2} \left( \partial_r^2  +
\left[ {d_a\over 2r} - |y|^2 \right] \partial_r \right. \cr  &  \left. + { d_a
(d_a-4) \over 16r^2}- {d_a\over 4r} |y|^2 + {|y|^4\over 4}   \right) . }$$
Each of the terms appearing on the right is of order $1/r^2$, and so none can a
priori be neglected. In a similar way, the rest of the terms in the Hamiltonian
modify the long distance behavior. This includes the $O(y^4)$ terms in the
potential which we recall is of the form, $V \sim r^2|y|^2/2 + O(y^4)$. Since
we need to include the $O(y^4)$ terms, this approximation is not one-loop in
the usual sense. We will need to sum up all the corrections to free
propagation, which are of order $1/r^2$, and surprisingly, they all cancel. The
remaining index computation then involves free particle propagation on the
moduli space which is $ {\bf R}^{(d_c-2)} /{\bf Z}_2$. From this computation,
we will recover the needed corrections to the index.

The construction that we shall describe is to be contrasted with the kind of
effective action for the light modes that has been obtained using
large-distance low-velocity expansions \refs{\rBFSS, \rDKPS}. After integrating
out the massive modes in a one-loop approximation, the leading correction to
the effective Lagrangian at large $r$ is a term of order $\sim v^4/r^7$ for the
case of the nine-dimensional model. The connection between that approach, and
the computations that we shall describe does not seem transparent. It seems
possible that exploring the connection in detail will give insight, and dare we
hope a proof, of the desired non-renormalization theorem for the $F^4$ term.
Computing the leading corrections to the three and five-dimensional models also
seems an interesting question. Since the amount of supersymmetry is reduced, we
might suspect that there is a correction to the metric on the moduli space.
However, in constructing the propagator using this approach, we do not find any
fundamental difference between the three cases.

Let us start by discussing the form of the various operators that we need to
study in special
coordinates. Without rotating to a convenient set of coordinates, it will be
very difficult to say anything about the structure of the partition function.
We can rotate our coordinates, $x$, into a convenient basis by using a
combination of gauge and flavor symmetries. So, we can choose a basis,
$$x=k\Lambda q,$$
with $k\in SO(3)$, $q\in SO(d)$ and $\Lambda$ the following $3\times d$ matrix,
\eqn\newcoord{ \Lambda = \left (\matrix{
r & 0 &  \cdots & & & 0 \cr
0 & y^2_2 & \cdots &  & & y^d_2 \cr
0 & y^2_3 & \cdots &  & & y^d_3 \cr
}\right ). }
We have set $\Lambda_1^1 = r$, and let $y$ denote the remaining $2\times (d-1)$
matrix, with $\Lambda^i_A=y^i_A, $ for $i,A > 1$. The reason for this choice is
that the flat directions are now at the locus, $y=0$. Note that the choice of
$k$ and $q$ is not unique here. So the mapping,
$$m(k,\Lambda,q)\, \r \, k\Lambda q,$$
projects $SO(3)\times {\bf R}^{2d-1}\times SO(d)$ onto our space of matrices,
$x$, but is clearly not one-to-one. The fibers of the map, $m$, are
non-trivial. Any function that depends on $x$ can then be lifted to a function
in the product space, which constant under those transformations of $ k,
\Lambda, q$ that leave $x$ invariant. Most of our computations will focus on
the neighborhood, $N_F$, of a flat point given by $|y|^2 < 1$, and $r$ tending
to $\infty$. This is the region that contributes to \boundary.

We need to write the Hamiltonian in terms of these new coordinates, which
include the $d+1$ angular variables parametrizing the $d+2$-dimensional flat
directions. The kinetic terms in the Hamiltonian can be determined by computing
the Laplacian for the metric associated to this coordinate choice. Let us
recall that given a metric, $g$, the Laplacian is given by:
$$ \Delta = - {1\over \sqrt{|g|}} \, T_i \left( \sqrt{|g|} g^{ij} \right) T_j,
$$
where $|g|$ is $ \det (g)$, and the $T_i$ are a basis of vector-fields for some
coordinate system. First, we need to make a choice of basis of vector-fields.
For the coordinates, \newcoord, it is natural to have $\{ {\partial\over
\partial r}, {\partial\over \partial y^i_B} \}_{2\leq i\leq d,B>1}$ as part of
the basis. We need $d+1$ additional vector-fields. From the left $SO(3)$, we
can choose the two vector-fields, $\{X_2,X_3\}$ which are associated to the two
$SO(3)$ generators,
$$ \left (\matrix{
0 & 1 & 0 \cr
-1 & 0 & 0 \cr
0 & 0 & 0 \cr
}\right) \, , \, \left (\matrix{
0 & 0 & 1 \cr
0 & 0 & 0 \cr
-1 & 0 & 0 \cr
}\right)  $$
respectively. Similarly, we can add the $d-1$ vector-fields $\{V_j\}_{j>1}$,
associated to the right $SO(d)$ generators, $z(j)$. The matrix $z(j)$ is a
$d\times d$ anti-symmetric matrix with only one positive entry, $z(j)_{1j} =
1$. Our total basis is then composed of the subset of tangent vectors to the
product space, $SO(3)\times {\bf R}^{2d-1}\times SO(d)$, given by
$\{X_2,X_3\}\cup\{ {\partial\over \partial r}, {\partial\over \partial y^i_B}
\}_{2\leq i\leq d,B>1}\cup
\{V_j\}_{j>1}.$

We need to determine the metric, $g$, for this coordinate choice. The set of
vector-fields, $\{ {\partial\over \partial r}, {\partial\over \partial y^i_B}
\}$ are orthonormal and orthogonal to the rest of the basis. The rest of the
basis have inner products,

$$(X_j,X_k) = r^2\delta_{jk} + (yy^t)_{jk}, $$
$$(V_j,V_k) = r^2\delta_{jk} + (y^ty)_{jk}, $$
and,
$$(X_j,V_k) = 2r y_{jk}.$$
These inner products can be determined by pushing forward the vector-fields
under $m$, and computing the resulting norms. Now the metric can be written as
a direct sum of two metrics, $g=g' \oplus g''$, where $g''$ is the identity
matrix for the coordinates corresponding to $(r,y)$. The interesting part of
the metric is the part for the angular variables. So, let us write, $g' = r^2 I
+ K$, where $K$ is determined from the above inner products. Then $(g')^{-1} =
I/r^2 - K/r^4 + \ldots$, where the omitted terms are suppressed by more powers
of $r$. To compute the Laplacian, we need:

$$ \eqalign{ {\rm log det} (g) = {\rm log det} (g') &= {\rm tr log}(r^2 I) +
{\rm tr log}(I+K/r^2)\cr &= 2(d+1) \log(r) + \tr (K/r^2)-\tr (K^2/2r^4) +
\ldots \cr &= 2(d+1)\log(r) - 2|y|^2/r^2 +\ldots,}$$
where omitted terms are again of lower order. Finally, this allows us to write
down an expression for the Hamiltonian in these special coordinates at
$(k,\Lambda,q),$
\eqn\hamil{ \eqalign{2H =& -{\partial^2\over\partial r^2} - {(d+1)\over r}
{\partial\over \partial r} + \Delta_y
+ {2 y^j_B \over r^2}  {\partial\over \partial y^j_B}
- {1\over r^2} (\sum_{j>1}X_j^2 + \sum_{j>1}V_j^2)  + \cr & r^2|y|^2 +
\sum_{i>j>1}(y^i_2y^j_3-y^i_3y^j_2)^2 +
k_{AM}\Lambda_M^s q_{sj}Y^j_A + \ldots, }}
where $\Delta_y$ is the Laplacian in the $y$ variables. We have written $H_F$
as
$k_{AM}\Lambda_M^s q_{sj}Y^j_A$, where $Y^j_A = i
\g^j_{\a\b}f_{ABC}\p_{B\a}\p_{C\b},$ in the nine-dimensional model, and
analogous expressions for the other cases. The omitted terms are all of order
$O(1/r^3)$ or smaller.

\subsec{Inverting the Hamiltonian}

To invert the Hamiltonian, let us focus first on the
harmonic oscillator term,
$H_m = \Delta_y + r^2|y|^2,$ with eigenvalues $2(d-1 + n)r,$ where $n$ a
non-negative integer.  If there were no cancelling fermion term, this
oscillator term would immediately guarantee a potential linearly increasing
with $r$, and therefore, a discrete spectrum.  In order to see
the cancelling fermion term, we write
$H_F = rY_r + H_F',$ where $ Y_r = k_{A1}q_{1j}Y^j_A,$
and $H_F'$ consists simply of all those terms in $H_F$ without a
$\Lambda^1_1=r$
factor.  It is an easy task to show that the eigenvalues of $Y_r$
range from $-2(d-1)$ to $2(d-1)$. In particular, when we consider wavefunctions
for which the transverse bosons and fermions are simultaneously in their ground
states, all terms of order $r$ cancel, and we are left with only terms of order
$1/r^2$ to worry about in the Hamiltonian. On any state with excited
oscillators, we see that $H$ has lowest eigenvalue of order $r$.

Our construction of $W$ will use the nice factorization of the wavefunctions in
terms of their behavior in the massive directions, and their behavior along the
flat directions. We can think of $H$ as a block $2\times 2$ matrix, with
respect to this decomposition, where the $H_{11}$ piece corresponds to the
terms in $H$ which take the ground state for the oscillators back to itself.
The piece $H_{22}$ contains terms in $H$ which send the state with one excited
massive boson to itself, and the off-diagonal terms act in a similar way. As a
first guess, one might try (and we did) to construct $W$ as a perturbation of
some nice approximation, $W_1$, to $H_{11}^{-1}\oplus H_{22}^{-1}$. Because of
the large eigenvalue of the first excited state of $H_m+rY_r$, almost any
construction should give a good $H_{22}^{-1}.$ Inverting $H_{11}$ is more
problematic but not excessively so.

In a standard perturbative approach, we consider $HW_1 = I - E_1.$ Here, $E_1$
will include, for example, such terms
as $H_{12}W_1.$  As a next step, set
$W_2 = W_1 + W_1E_1,$ with error $E_2 = -E_1^2.$
One could iterate this construction to construct $W = W_n$
for some large $n$, if the errors were getting significantly smaller
each time. For example if each $E_k$ had a kernel $e_k(x,x')$ which was
bounded by $(r(x) + r(x'))^{-k},$ this would lead us to the desired $W$ after
some number of iterations. With this in mind, it becomes clear
what terms must be included in our initial approximation, $W_1$. Because any
approximate $H_{11}^{-1}$ will have upper bound of size $r^2$,
we see that we cannot discard any term of size bigger than or equal to
$O(1/r^2)$ in $H$ which either acts on the ground state, or maps an excited
state to the ground state. For these terms lead to errors which are not
decreasing under iteration.  Actually,
this is an overstatement. If a term $B$ maps us, for example,
from the lowest state into a higher state we see that the error will enter as
$H_{22}^{-1}BH_{11}^{-1}.$ Because $H_{22}^{-1}$ is bounded above
by $(r(x) + r(x'))^{-1}$, this term will be decreasing in $r$ if,
for example,  $B$ is $O(1/r^2)$.  With these
remarks to guide us, we must now return to analyze $H$,
treating as lower order only terms which are $O(1/r^3)$ if they map
the ground state into itself, or are $O(1/r^2)$ if they mix the
ground state with excited states.  These lower order terms
cannot be neglected altogether, but they simply enter as
higher terms in our iterative construction of $W$.

The gauge constraints provide some further simplication in our computations.
First we can `move' to the identity element $k=1$ in our product space,
$SO(3)\times {\bf R}^{2d-1} \times SO(d)$ with coordinates $ (k,\Lambda, q)$,
by a gauge rotation. From now on, we will restrict our discussion to the $k=1$
subspace. Further, setting two of the gauge constraints to zero allows us to
replace differentiation by $X_j$, which generates gauge transformations on the
bosons, by multiplication by the fermion bilinear which generates gauge
transformations on the fermions. For example, in the nine-dimensional case, the
fermion bilinear is given by $Q_j := -{1\over 2} \p_{1s}\p_{js}.$  There are
similar expressions for the other cases. The remaining gauge constraint, $C_1$
of the three constraints in \constraints, generates a $U(1)$ subgroup which
acts on $y$. With these considerations in hand, let us turn to the task of
getting rid of the massive modes, and obtaining an effective Hamiltonian on the
flat directions which we can invert to compute \boundary.

\subsec{Constructing the effective Hamiltonian}

Let us begin by computing the total contribution of terms that map the ground
state to itself. Each of these terms will give rise to an interaction in the
effective Hamiltonian of the form, $m/ r^2$, for some $m$, in a manner
described in the beginning of this section. The ground state is of the form
$s(q) r^{(d-1)/2} e^{-r|y|^2/2}$, where $s(q)$, the fermion ground state, is
actually a section of a
bundle determined by the lowest eigenvalue of the fermion term $Y_r$, since
$s(q)$ must satisfy the equation:
$$ \{ Y_r(q) + 2(d-1) \} s(q) = 0.$$
Therefore, the ground state depends non-trivially on the right angular
coordinates, $q$. We will examine the structure of the fermion ground state in
some detail, shortly. Note that the ground state is invariant under the
remaining $U(1)$ subgroup of the gauge group. As a first approximation, a
general state that we need to consider is a product of the ground state with a
wavefunction, $f(r,q)$, along the flat directions. Let us record the various
$m^i/r^2$ contributions to the effective Hamiltonian which acts on $f$, where
$m^3, m^5$ and  $m^9$ denote the values of the contributions for the three,
five and nine-dimensional models, respectively. First, by definition, the term
$H_m + rY_r$ vanishes on the ground state.

It is convenient to rewrite $y$ and $ {\partial\over \partial y}$ in terms of
standard annihilation and creation operators:
$$ \eqalign{ y &= {1\over \sqrt{2r}} ( a+a^\dagger ), \cr
             {\partial\over \partial y} &= \sqrt{r \over 2} (a- a^\dagger),}$$
where $ [a,a^\dagger ] =1$.
Now we can evaluate the contribution of the derivative terms in $H$ acting on
the bosonic ground state,
 $$ \vac = \left( {r\over \pi} \right)^{(d-1)/2} e^{-r|y|^2/2}.$$ The term in
\hamil\ with one radial derivative gives,
 $$ \eqalign{ \lvac \partial_r \vac &= \lvac \left( {d-1\over 2r} - {y^2\over
2} \right) \vac \cr & =0.} $$
 The term with two radial derivatives then gives,
 $$ \eqalign{ \lvac \partial_r^2 \vac &= {1\over r^2} \left[ ({d-1\over
2})({d-3\over 2}) - 2({d-1\over 2})^2 \right] + {1\over 4}\lvac  |y|^2
|y|^2\vac \cr
 &= {1-d \over 4 r^2}.} $$
 Let us label this contribution, $m_r$, then $m^3_r=1/4$, $m^5_r=1/2$, and
$m^9_r=1$. After noting that we can replace the $ X_i^2$ terms in \hamil\ by
the action of the fermion bilinears, $Q_i^2$, which are just matrices, we see
that there are two remaining derivative terms in the Hamiltonian. The first is
the $ y {\partial \over \partial y}$ term and it is relatively easy to analyze,
 $$ \eqalign{ \lvac y {\partial \over \partial y} \vac &= -{1\over 2} \lvac a
a^\dagger \vac \cr &= (1-d).}$$
Let us call this contribution, $m_y$, where $m^d_y =1-d $. The sign of this
term is critical since it is the only term that maps the ground state to the
ground state, and gives a large negative contribution to the net $m$.

The last derivative term comes from the action of $ V_i^2$ on the ground state,
where the $V_i$ generate certain flavor rotations. The $q$-dependent fermion
oscillator ground state is the only part of the full ground state wavefunction
that can give a non-vanishing contribution in this case. So, let us now
describe the ground state of the fermions in some detail.  The real massive
fermions, $ \{ \p_{2\a}, \p_{3\a} \} $, where $\a$ runs from $1$ to $n=4,8$ and
$16$ for the three, five and nine-dimensional cases,  can be arranged into
annihilation and creation operators:
$$ \eqalign{ b_\a &= {1\over \sqrt{2} } ( \p_{2\a} + i \p_{3\a} ), \cr
             b^\dagger_\a &={1\over \sqrt{2} } (\p_{2\a} - i \p_{3\a}). \cr} $$
The operators $ b,b^\dagger$ obey the anti-commutation relation, $ \{ b_\a,
b^\dagger_\b \} = \delta_{\a\b}$. Can we construct a fermion state in the
kernel of $ Y_r + 2(d-1)$? For the moment, let us pick $q=1$. In this case,
$Y_r = 2i \g^1_{\a\b} \p_{2\a} \p_{3\b} $, and we can pick $\g^1$ to be the
diagonal element of the Clifford algebra:
$$ \left (\matrix{
1 & 0 \cr
0 & -1  \cr
}\right). $$
With this choice, $Y_r = 2 \sum_{\a=1}^{n/2}{ \left( b^\dagger_\a b_\a
-b^\dagger_{n/2+\a} b_{n/2+\a} \right) }$. Let us choose a fermionic ground
state $\vac_F$ which satisfies, $ b_\a \vac_F = b_{n/2+\a}^\dagger \vac_F= 0$
for $\a=1,\ldots, n/2.$ This vacuum is then in the kernel of $ Y_r + 2(d-1)$ at
the point $q=1$. We should point out that, when restricted to the
$(d-1)$-sphere which is the  SO(d) orbit of a flat point, the ground state
takes values in a flat vector bundle, which is therefore trivial.  This means
that there is a globally defined fermion ground state wavefunction.

To understand the fermion ground state for arbitrary $q$ and, in particular, to
understand the action of $V_i^2$ on the ground state, we shall study the
equivalent problem of its action on the operator which acts as
projection onto the kernel of $ Y_r + 2(d-1)$. This operator is given by,
\eqn\project{P(q) = {1\over 2\pi i} \oint_{\Gamma} dz {1\over z - Y_r(q)}, }
using a contour, $\Gamma$, which we will take to be a small loop enclosing
$-2(d-1)$. This construction of $P(q)$ is readily seen
to be correct by diagonalizing $Y_r(q)$, and computing the corresponding
contour integral.

 Here we should clarify that we
are really interested in the action of $P(q)\sum_iV_i^2$; the remainder is
$O(1/r^2)$ and does not take the groundstate to itself. These terms will
therefore
be of interest only as perturbative corrections. The operator
$P(q)\sum_iV_i^2$  should
act as a scalar $2m_q$ on an appropriately chosen basis of the ground
state. Our task is to compute the scalar. Now,
$P(q)$ is given by conjugating the projection operator at $q=1$ with
a nonconstant orthogonal matrix, whose columns give a basis of
the ground state. We can write this as,
$P(q) = O(q)P(1)O(q)^{t}.$ Then $\sum_i V_i^2$ acting on $P(q)$
has terms where $O(q)$ or $O(q)^t$ are twice-differentiated and terms where
each is differentiated once.  It is easy to show that this last term is
annihilated by $P(q)$, and hence is not germane to this calculation.
If the remaining term is a multiple of $P(q)$, then the multiple
will be $4m_q$, as both $O(q)$ and its conjugate should contribute
a factor of $2m_q$ when differentiated.
We preface the computation by commenting on how to compute the derivative
of $Y_r(q)$. It is enough to compute $V_i q_{1j}$.
By a change of coordinates, we then only need to compute,
$V_iq_{1j}(1).$ This is given by:
$$V_iq_{1j} = \frac{d}{dt}(q e^{tV_i})_{1j} \Bigr|_{t=0} = (q V_i)_{1j}.$$
Evaluating at $q=1$ gives,
$$V_iq_{1j}(1) = (V_i)_{1j} = \delta_{ij}.$$
Now we compute:
$$ \eqalign{ \sum_i V_i^2 P(q) =& \sum_i {2\over 2\pi i} \int {1\over z -
Y_r(q)}V_iY_r{1\over z - Y_r(q)}V_iY_r{1\over z - Y_r(q)}dz
+ \cr &
\sum_i {1\over 2\pi i} \int {1\over z - Y_r(q)}V_i^2Y_r{1\over z - Y_r(q)}dz,
}$$
where $Y_r$ is an eigenfunction of the Laplacian $\sum_i V_i^2 $. Hence the
second integrand
is a simple double pole and therefore integrates to zero.
In order to compute the remaining term, it is enough to make a change
of coordinates equivalent to taking $q=1$. Then
it is easy to see that $V_i(Y_r)$ takes the $-2(d-1)$ eigenspace of $Y_r$
to the $4-2(d-1)$ eigenspace. This gives,
$$ \eqalign{ & P(q)\sum_i  {2\over 2\pi i} \int {1\over z -
Y_r(q)}V_iY_r{1\over z - Y_r(q)}V_iY_r{1\over z - Y_r(q)}dz = \cr &
\sum_i \hbox{diag}(V_iY_r)^2 {2\over 2\pi i}\int {1\over(z+2(d-1))}{1\over
(z+2(d-1) -4)}{1\over (z+2(d-1))}dz, }$$
where $\hbox{diag}(V_iY_r)^2 = P(q)(V_iY_r)^2V(q).$
Integrating gives $-\sum_i \hbox{diag}(V_iY_r)^2/8.$

We can compute this at $q=1$ where,  $(V_iY_r) = 2i
\g^i_{\a\b}\p_{2\a}\p_{3\b},$
and
$\sum_i(V_iY_r)^2/8 = -4\g^i_{\a\b} \p_{2\a} \p_{3\b}
\g^i_{\a'\b'}\p_{2\a'}\p_{3\b'}/8.$
The terms which survive $P(q)$ are,
$$ \eqalign{ -4 \g^i_{\a\b}\p_{2\a}\p_{3\b}\g^i_{\a\b}\p_{2\a}\p_{3\b}/8
-4\g^i_{\a\b}\p_{2\a}\p_{3\b}\g^i_{\b\a}\p_{2\b}\p_{3\a}/8 &=
\g^i_{\a\b}\g^i_{\a\b}/4 \cr
&= 4(d-1).}$$
So, we finally get that $m^3_q = 1/2, m^5_q = 2$ and $m^9_q=8.$

With the derivative terms out of the way, we can consider the two remaining
operators in the Hamiltonian that map the ground state to the ground state. The
first is the $O(y^4)$ term in the potential. This gives,
$$ \eqalign{ \lvac \sum_{i>j} (y_2^i y_3^j - y_3^i y_2^j)^2 \vac &=  \lvac
2\sum_{i>j} (y_2^i y_3^j)^2 \vac \cr &= {1\over 4 r^2} (d-1)(d-2).}$$
Calling this contribution, $m_V$, we have $m^3_V=1/4, m^5_V=3/2,$ and
$m^9_V=7$. The last contribution comes from the kinetic term for the two gauge
rotation generators, ${X_2, X_3}$. Setting the gauge constraints to zero, we
can replace the angular Laplacian, $ X_i^2$, by the operator quartic in
fermions:
$$ \sum_{\a=1}^{n}{ (\p_{1\a} \p_{2\a})^2 + (\p_{1\a} \p_{3\a})^2.} $$
The only terms that map the ground state to the ground state are those
proportional to the identity. A quick calculation gives the numerical value,
$n/2$. Calling this contribution, $m_f$, we note that $m^3_f=1, m^5_f=2$ and
$m^9_f=4$.

If we were to stop at this point, and just consider these diagonal
contributions to the effective Hamiltonian, a quick check would show that the
net $m$ is non-vanishing. There would therefore be a non-trivial $1/r^2$
interaction in the effective theory. Fortunately, we are not quite finished.
First, we can shift the coefficient of the $ \partial_r$ term in \hamil, and
generate a new $O(1/r^2)$ term by redefining our wavefunctions. The reason this
is useful is that on choosing an appropriate  coefficient for  $\partial_r$, we
can combine the radial derivatives with the $V_i$ angular derivatives to obtain
a Laplacian for flat space, together with some $1/r^2$ interaction. This way,
we
only need to deal with Euclidean coordinates, rather than the messier
angular coordinates. We want to shift the coefficient of the $ \partial_r$ term
from $d+1$ to $d-1$, so we will end up with a free particle Hamiltonian on the
$d$-dimensional moduli space, together with interactions. To do so, we note
that:
$$ \eqalign{  - \left( \partial_r^2 + {d+1\over r} \partial_r \right) &=
 - \left( (\partial_r + {1\over r})^2 + {d-1\over r} \partial_r  \right) \cr &
=- \left( (\partial_r + {1\over r})^2 + {d-1\over r} (\partial_r+{1\over r})
-{d-1\over
r^2} \right) \cr &= - {1\over r} \left( \partial_r^2 + {d-1\over r} \partial_r
-{d-1\over r^2} \right)r ,}$$
where we now redefine our wavefunctions, $f(r,q) = {1\over r} \tilde{f}(r,q)$.
This
gives us a new contribution to $m$, say $m_c={d-1\over2}$.

So far, we have found that $H$ acts on the ground state and first excited state
in the following way,
$$ H =\left (\matrix{
{1\over 2}\Delta_d + m^d_T / r^2 & b^t \cr
b & H_{22}  \cr
}\right).$$
Here, $m_T = m_c+m_f+m_V+m_q+m_y+m_r$ is the total effective interaction that
we have found so far, where $m^3_T=1, m^5_T=4 $ and $m^9_T=16$. Lower order
terms in $H_{11}$ have been omitted. Let us turn to the form of $b$, which may
change the effective interaction. For example, the terms in $b$ acting on the
ground
state  of order $O(r^{-1/2})$ cannot be neglected. Our initial choice of
$W_1$ is then not diagonal to the requisite order in the basis that we have
been using. Does $b$ contain terms of the right order?

It is not hard to check that the only terms that map the ground state to the
first excited state, which are of the requisite order are those involving
$H_F'$. This follows from noting that $H_F'$ is proportional to $y$ which is,
$\sim a^\dagger/\sqrt{r} $ acting on the ground state. Now we see that we can
significantly lower our energy if we define a new ground state $ \vac' =
(I-{b\over 2r}) s(q) \vac$. The factor of $2r$ is chosen because acting with
$b$ on a state raises its eigenvalue under $H_m + rY_r$ by $2r$. So, $H\vac' =
({1\over 2}\Delta_d + m^d_T/r^2 - b^t b/2r) s(q)\vac$ plus lower order terms.
The final
contribution to $m$ from $b^tb$ is computed in the same way as the other
contributions, and we find that $m^d_b = -m^d_T$.

Wonderfully, the $O(1/r^2)$ terms sum to zero! From these considerations, we
see that we are reduced to a free-particle calculation. Our choice for an
approximation $W_{11}$  to $H_{11}^{-1}$ is then particularly simple: we can
just take the free-particle propagator,
$$ W_{11} = \int{ {d^dk \over (2\pi)^d} \, { e^{i k \cdot (x-x')} \over k^2}}.
$$

\subsec{Evaluating the boundary contribution}

Now that we reached the point where we have a nice simple form for $W_{11}$, we
note that the remainder of the perturbation construction is standard. We will
not belabor the reader with the details of this expansion, but just provide
some relevant comments. As we observed before, any reasonable construction
yields a good approximate $H_{22}^{-1}$, with a nice error bound. The
construction of $W$, which is perturbative in $1/r$, then follows the outline
that we have described earlier in this section. The contribution of the
$W_{11}$ term is non-vanishing, but it is clear after extensive, arduous but
standard computations, which involve checking powers of $r$, that any trace
involving the rest of $W$ will bring in more powers of $r^{-1}$
than appear in the $W_{11}$ term. These terms will therefore not contribute to
the boundary term \boundary, in the limit where $r \r \infty$. We can now
restrict ourselves to the free-particle Green's function, $W_{11}$. The
remaining calculation is simple. We have a free particle propagating on ${\bf
R}^d / {\bf Z}_2$. The ${\bf Z}_2$ identification comes from the Weyl group
action on the Cartan of the gauge group $SU(2)$. Let us take $x$ as coordinates
for ${\bf R}^d$. The ${\bf Z}_2$ action acts as parity, sending $x \r -x$. It
also sends the free fermions, $\p_{1\a}$, where $\a=1,\ldots, 2(d-1)$, to minus
themselves. The Hilbert space of gauge-invariant wavefunctions is given by:
$$ \{ f_0(x), f_1(x) \p_{1\a}, f_2(x) \p_{1\a}\p_{1\b}, \ldots \}.$$
Each function, $f_k$, has parity $(-1)^k$. All we need to do is compute the
boundary term,
$$ -{1\over 2}\int_{N_{F}(r)} \tr e_n(-1)^FQW_{11}, $$
for this system. This becomes an integral over the boundary of the
$d$-dimensional moduli space:

$$ \eqalign{-{1\over 2}\int_{S^{d-1}(r)} \tr e_n(-1)^{F}QW_{11} &= - {1\over
\vol(S^{d-1})2 (d-2)} \int_{S^{d-1}(r)} \tr e_n (-1)^F Q {1\over |x-x'|^{d-2}
}\Bigr|_{x'=-x} \cr &= -1/4. \cr}$$

To close this computation, let us note that the lower bound on the asymptotic
behavior of $H$ is given by its lowest value on the modified lowest
wavefunctions, where the modification involved the off-diagonal $b$ term. The
lower bound is therefore the same as the bound for $\Delta_d$, up to terms of
order $O(1/r^3)$. As discussed in section three, this easily leads to the
claimed asymptotic lower bound for $H$.

To summarize: we have found a formula for the index that counts the net number
of $L^2$ ground states in certain quantum mechanical systems, where the
potential has flat directions. This involved a study of $L^2$ index theory for
a family of
non-Fredholm operators, which allowed us to show that the prescription we
presented actually computes the index. For the case of two-particle binding, we
have shown that there is a bound state for coincident zero-branes in type IIA
string theory. We have also found further evidence that there are no bound
states for two-branes twice wrapped on an $S^2$, and three-branes twice wrapped
on an $S^3$. Note that these models are only special points in the space of
theories obtained by deforming the zero-brane quantum mechanics.

The actual computation split into two parts. Computing the principal term
involved evaluating the integral \finalint. It would be interesting, and quite
non-trivial, to compute this integral for higher rank gauge groups. Even better
would be a method for avoiding this integration altogether. The second part of
the computation required a study of the propagator for the two particles when
they are far apart. Surprisingly, after summing a variety of corrections, this
computation reduced to one involving a free particle moving on the moduli
space. Undoubtedly, there is a fundamental reason for this simplification, and
finding it may also shed light on whether the $F^4$ term in the effective
zero-brane Hamiltonian is protected from corrections. It seems likely that
there will be an analogous reduction to a free particle calcuation for other
gauge groups. As a further comment, note that if we had studied a system with
gauge group $U(1)$ and some charged matter, there would have been no boundary
correction, as in the case involving H-monopoles \rhmon.

The sort of decay estimates that we described can probably be used to get a
handle on the structure of the ground state wavefunction. What is needed is an
upper bound on how fast the wavefunction can decay along the flat directions.
They may also lead to a vanishing theorem showing that all ground states in
these systems must have a definite fermion number. The index would no longer be
just an index, but would then count the total number of ground states. This
would allow us to conclude that the zero-brane bound state is unique. Finally,
systems involving marginal binding of branes with different dimensions can now
be analyzed in much the same way.

\bigbreak\bigskip\bigskip\centerline{{\bf Acknowledgements}}\nobreak

It is a pleasure to thank G. Jungman, L. Susskind and E. Witten for helpful
discussions. The
work of S.S. is supported by NSF grant DMS--9627351, while that of M.S. by NSF
grant DMS-9505040.

\listrefs

\end